\documentclass[12pt,a4paper,final]{iopart}

\usepackage{iopams}  
\usepackage{graphicx}
\usepackage[breaklinks=true,colorlinks=true,linkcolor=blue,urlcolor=blue,citecolor=blue]{hyperref}

\newcommand{\iden}{\hat{\mathbb{I}}}

\newcommand{\ketbra}[1]{\left|#1\right>\hspace{-4pt}\left<#1\right|}
\newcommand{\ketbrat}[2]{\left|#1\right>\hspace{-4pt}\left<#2\right|}

\begin{document}

\title[A randomized benchmarking suite for mid-circuit measurements]{A randomized benchmarking suite for mid-circuit measurements}

\author{L.~C.~G.~Govia}
\address{IBM Quantum, IBM Almaden Research Center, San Jose, CA 95120, USA}
\ead{lcggovia@ibm.com}
\author{P.~Jurcevic}
\address{IBM Quantum, IBM T.~J.~Watson Research Center, Yorktown Heights, NY 10598, USA}
\author{C.~J.~Wood}
\address{IBM Quantum, IBM T.~J.~Watson Research Center, Yorktown Heights, NY 10598, USA}
\author{N. Kanazawa}
\address{IBM Quantum, IBM Research Tokyo, 19-21 Nihonbashi Hakozaki-cho, Chuo-ku, Tokyo, 103-8510, Japan}
\author{S.~T.~Merkel}
\address{IBM Quantum, IBM T.~J.~Watson Research Center, Yorktown Heights, NY 10598, USA}
\author{D.~C.~McKay}
\address{IBM Quantum, IBM T.~J.~Watson Research Center, Yorktown Heights, NY 10598, USA}

\begin{abstract}
Mid-circuit measurements are a key component in many quantum information computing protocols, including quantum error correction, fault-tolerant logical operations, and measurement based quantum computing. As such, techniques to quickly and efficiently characterize or benchmark their performance are of great interest. Beyond the measured qubit, it is also relevant to determine what, if any, impact mid-circuit measurement has on adjacent, unmeasured, spectator qubits. Here, we present a mid-circuit measurement benchmarking suite developed from the ubiquitous paradigm of randomized benchmarking. We show how our benchmarking suite can be used to both detect as well as quantify errors on both measured and spectator qubits, including measurement-induced errors on spectator qubits and entangling errors between measured and spectator qubits. We demonstrate the scalability of our suite by simultaneously characterizing mid-circuit measurement on multiple qubits from an IBM Quantum Falcon device, and support our experimental results with numerical simulations. Further, using a mid-circuit measurement tomography protocol we establish the nature of the errors identified by our benchmarking suite.
\end{abstract}

\maketitle

\section{Introduction}
\label{sec:intro}

Steady progress towards quantum error correction and fault tolerant quantum computing has led to recent experimental demonstrations of small quantum error correcting codes \cite{Egan21,Chen21,Postler2021,Satzinger21,Ryan-Anderson21,Abobeih21,Chen21IBM,Krinner21,Zhao21,Sundaresan22}. An essential component to most implementations of quantum error correction is the ability to repeatedly measure stabilizers of the code, often achieved via a stabilizer check circuit that encodes the outcome of the stabilizer measurement into the state of an ancilla qubit. Whether via an ancilla or measured directly \cite{Riste13,Livingston22}, stabilizer checks require fast and accurate mid-circuit measurement. Thus, characterizing and benchmarking mid-circuit measurement is a key capability for the development and execution of fault tolerant quantum computing.

Going beyond the typically measured state-assignment fidelity, quantum detector tomography \cite{Luis99,Fiurasek01,DAriano04,Lundeen09,YChen19} can be used to characterize terminal measurements in terms of a positive operator-valued measure (POVM). However, for the characterization of mid-circuit measurements a POVM description is insufficient. In this case full characterization of the measurement action leading to each outcome is described by a quantum channel, and hence process tomography is required \cite{qilgst,Pereira21,Pereira22}. While one can imagine extensions of these protocols to small stabilizer check circuits, the exponential resource scaling of process tomography make such characterization approaches impractical to deploy for larger quantum codes. While providing less detailed information about the measurement operation, there is a need for a scalable benchmark that can quickly assess the performance of mid-ciruit measurement, and how it impacts not only the measured qubit but also those qubits connected to it.

Here, we introduce the mid-circuit measurement randomized benchmarking (mcm-rb) suite as one such benchmark. Building off the well-studied family of randomized benchmarking (RB) protocols \cite{Helsen20,Knill08,Magesan11,Magesan12,Magesan12b,Gambetta12,Wallman15,Wallman16,Cross16,Wood18,McKay19,Helsen19,Proctor19,Erhard19,McKay20,Proctor21,Morvan21}, the mcm-rb suite comprises a central protocol, {\tt mcm-rb}, that interleaves measurements of an ancilla qubit between the gates of Clifford RB performed on a distinct control qubit. The two other protocols of our suite are control experiments that replace either the measurement ({\tt delay-rb}) or the Clifford gates ({\tt mcm-rep}) with delays of equal time duration to the replaced operation. Comparison between the decay curves of {\tt mcm-rb} with the control experiments allow for the identification, and in some cases quantification, of the error induced by measurement on the control and ancilla qubits under the standard RB assumptions (i.e.~Markovianity \cite{Knill08}). The mcm-rb suite is highly scalable, as it can be applied to many control and ancilla qubits at once to simultaneously benchmark measurement induced error, including measurement cross-talk.

This manuscript is organized as follows. In section \ref{sec:mcmrb} we describe the procedure of the mcm-rb suite, and in section \ref{sec:error} we discuss a classification of errors that the suite can detect. In section \ref{sec:exp} we demonstrate the mcm-rb suite on an IBM Quantum device, and in section \ref{sec:sims} we present supporting numerical simulations. Finally, in section \ref{sec:Disc} we discuss a limitation and potential extension of the protocol, and in section \ref{sec:Conc} we make our concluding remarks. Further details of our experimental and numerical results can be found in the Appendices. While writing this manuscript, we became aware of Ref.~\cite{Gaebler21}, which demonstrates the {\tt mcm-rb} protocol, but not the rest of the suite, on a trapped ion system to study the impact of both mid-circuit measurement and reset.

\section{Mid-circuit Measurement Randomized Benchmarking Suite}
\label{sec:mcmrb}

\begin{figure*}[t]
    \centering
    \includegraphics[width=\columnwidth]{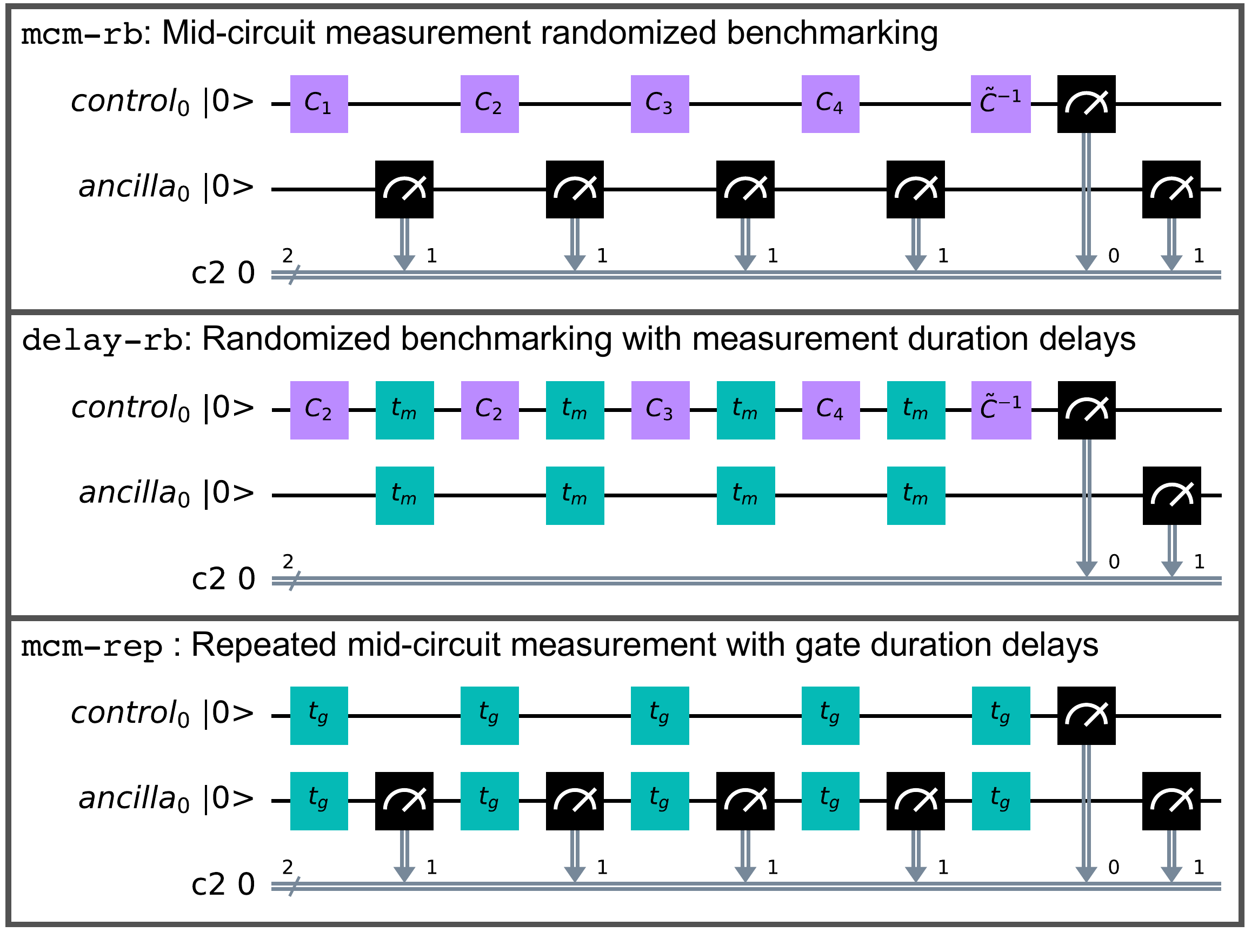}
    \caption{{\bf Example circuits of the mcm-rb suite protocols.} Sequence length $N_i = 4$ for all three protocols. (Upper panel) {\tt mcm-rb} circuit with Clifford gates on the control qubit interleaved by measurements on the ancilla. (Middle panel) {\tt delay-rb} circuit where Clifford gates on the control qubit are interleaved wiht delays of duration $t_m$, the length of an ancilla measurement. (Lower panel) {\tt mcm-rep} with repeated measurements on the ancilla qubit interleaved by delays of duration $t_g$, the length of a control qubit Clifford gate.}
    \label{fig:circuits}
\end{figure*}

The mcm-rb suite is defined by the set of RB-style protocols ({\tt mcm-rb}, {\tt delay-rb}, {\tt mcm-rep}), for which example circuits are shown in Fig.~\ref{fig:circuits}. The first protocol, {\tt mcm-rb}, interleaves ancilla-qubit mid-circuit measurements between Clifford gates performed on the control qubit. Similar to simultaneous RB \cite{Gambetta12}, it performs a single-subsystem twirl on the potentially two-qubit error induced by ancilla measurement. The second protocol, {\tt delay-rb}, is analogous to interleaved RB (IRB) \cite{Magesan12b} on the control qubit, with the interleaved gate a noisy identity corresponding to a delay of equal duration to the ancilla-qubit measurement.

Together, these two protocols form an IRB procedure designed to detect errors on the control qubit induced by the ancilla-qubit measurement. Though it contains an interleaved gate itself, {\tt delay-rb} is the reference sequence, and {\tt mcm-rb} is the interleaved error sequence. It is important to reference {\tt mcm-rb} by a sequence that contains interleaved delays in order to remove the trivial $T_1$ and $T_2$ decay of the control qubit during the potentially long measurement time, as this error may otherwise dominate other errors induced by the ancilla-qubit measurement. For example, for the experimental results of section \ref{sec:exp} the measurement time approaches $1~\mu$s, which should be compared to the much shorter gate time on the order of $10$s of ns and the several hundred $\mu$s $T_1$ and $T_2$ times.

The final protocol, {\tt mcm-rep}, is included to detect errors on the ancilla due to its own measurement, and is a modification of quantum non-demolition tests, e.g.~Ref.~\cite{Chen21IBM}. Delays of equal duration to the control-qubit Clifford gates are interleaved between repeated measurements to keep all three protocols of equal duration (for a given sequence length), and to detect measurement and logical basis misalignment. We discuss the latter point in more detail in section \ref{sec:exp}. Comparing {\tt mcm-rep} to {\tt mcm-rb} is also useful for detecting certain kinds of two-qubit errors that require one qubit to be excited, of which we show an example in sections \ref{sec:exp} and \ref{sec:sims}.

The mcm-rb suite is implemented similar to any RB-style protocol. A set of sequence lengths, $\{N_i\}_i$, is chosen, and for {\tt mcm-rb} and {\tt delay-rb} each sequence of length $N_i$ consists of $N_i$ random single-qubit Clifford gates on the control qubit interleaved by either measurements on the ancilla, or delays of equal duration, respectively. A final Clifford gate that is meant to invert the action of the previous $N_i$ Cliffords terminates every circuit. At each sequence length, many random Clifford circuits are executed. For {\tt mcm-rep} the circuit consists of $N_i$ ancilla measurements interleaved by delays of equal duration to the control qubit Clifford gates. In this manuscript we have chosen $N_{\max} = 150$.

For all three protocols, the outcomes of the mid-circuit measurements are discarded, and the ground state probabilities for all control and ancilla qubits at the end of each circuit is estimated from the terminal measurement. These probabilities are averaged over the random Clifford circuits and for each control or ancilla qubit the decay as a function of sequence length is fit to the exponential function $P_0 = A\alpha^{N_i} + B$. The RB-decay parameter $\alpha$ defines the error per Clifford/measurement for each qubit by EPC/M $= (1-\alpha)/2$, while the other fit parameters $A$ and $B$ account for system preparation and measurement error (SPAM). Under the standard assumptions of RB \cite{Knill08} {\tt mcm-rb} and {\tt delay-rb} will follow an exponential decay curve from which SPAM can be isolated from EPC, but this is not guaranteed for {\tt mcm-rep}. However, if the error on the measurement leads to monotonic decay then its EPM can be isolated from SPAM, though our fitting procedure makes the further (generically unnecessary) requirement that the decay is exponential. We have chosen not to Clifford twirl the measured ancilla qubits as doing so would require a final inverse gate that was conditional on the full history of measurement outcomes.

While Fig.~\ref{fig:circuits} shows as an example only a single control and ancilla qubit, the mcm-rb suite can be applied simultaneously to multiple control and ancilla qubits. This can be used, for example, to test the impact of measurement of a central ancilla on multiple control qubits, or to test the impact of the measurement of multiple ancilla qubits on a single control. In our experimental demonstrations of section \ref{sec:exp} we study both applications by simultaneously performing the mcm-rb suite across 12 control and 5 ancilla qubits on our device. While simultaneous Clifford gates on control qubits can introduce cross-talk error, since {\tt mcm-rb} and {\tt delay-rb} operate under the same control conditions with simultaneous gates they equivalently experience cross-talk. Thus, we can still use {\tt delay-rb} as the reference sequence to quantify the error induced by measurement in {\tt mcm-rb}.

\section{Error detection with mid-circuit measurement RB}
\label{sec:error}

In this section we demonstrate the capability of the mcm-rb suite ({\tt mcm-rb}, {\tt delay-rb}, {\tt mcm-rep}) to detect, and in many case estimate the magnitude of, errors induced by mid-circuit measurement on either the control or (measured) ancilla qubit. To do so, rather than focus on the effects of specific errors, we classify the distinct \emph{error signatures} that the mcm-rb suite's decay curves can exhibit, where each error signature can have more than one possible underlying physical error mechanism.

Error signatures are classified by the comparing the error per Clifford (EPC) of the control and error per measurement (EPM) of the ancilla for the three components of the mcm-rb suite. From here one, we denote the EPC and EPM by $\epsilon_{\nu}^{q}$, with $q\in\{c,a\}$ for control and ancilla respectively, and $\nu\in\{{\rm rb,del,rep}\}$ for {\tt mcm-rb}, {\tt delay-rb}, and {\tt mcm-rep} respectively. Table \ref{tab:esigs} outlines the error signatures we consider, and the expected relationships between the various EPCs and EPMs.

\begin{table}[!t]
    \centering
    \begin{tabular}{|l||l|}
    \hline
         {\bf Error Signature} & {\bf EPC/M}  \\
    \hline
    \hline
        No measurement & $\epsilon_{\nu}^{a} \approx 0~\forall \nu$  \\
        induced error & $\epsilon_{\rm rb}^{c} \approx \epsilon_{\rm del}^{c}$, $\epsilon_{\rm rep}^{c} \approx 0$  \\
    \hline
        Non-QND & $\epsilon_{\rm del}^{a} \approx 0$, $\epsilon_{\rm rb}^{a} > 0$, $\epsilon_{\rm rep}^{a} > 0$   \\
        measurement error & $\epsilon_{\rm rb}^{c} \approx \epsilon_{\rm del}^{c}$, $\epsilon_{\rm rep}^{c} \approx 0$  \\
    \hline
        Measurement induced  & $\epsilon_{\nu}^{a} \approx 0~\forall \nu$  \\
    control error & $\epsilon_{\rm rb}^{c} > \epsilon_{\rm del}^{c}$, $\epsilon_{\rm rep}^{c} \geq 0$  \\
    \hline
        Measurement induced  & $\epsilon_{\rm del}^{a} \approx 0$, $\epsilon_{\rm rb}^{a} \geq 0$, $\epsilon_{\rm rep}^{a} \geq 0$  \\ 
        2-qubit error & $\epsilon_{\rm rb}^{c} > \epsilon_{\rm del}^{c}$, $\epsilon_{\rm rep}^{c} \geq 0$  \\
    \hline
        RB cross-talk error & $\epsilon_{\rm rep}^{a} \approx 0$, $\epsilon_{\rm rb}^{a} > 0$, $\epsilon_{\rm del}^{a} > 0$  \\
        & $\epsilon_{\rm rb}^{c} \approx \epsilon_{\rm del}^{c}$, $\epsilon_{\rm rep}^{c} \approx 0$  \\
    \hline
    \end{tabular}
    \caption{Error signatures detected by the mcm-rb suite that we consider in this paper. For each error signature, the expected relationships between the EPCs and EPMs are shown. Each $\epsilon_{\nu}^{q}$ is an EPC/M with $q\in\{c,a\}$ for control and ancilla, and $\nu\in\{{\rm rb,del,rep}\}$ for {\tt mcm-rb}, {\tt delay-rb}, and {\tt mcm-rep}.}
    \label{tab:esigs}
\end{table}

It should be noted that multiple physical errors which result in distinct error signatures can occur simultaneously, such that the decay curves of the mcm-rb suite display a combination of the error signatures listed in Table \ref{tab:esigs}. In this case, one has to determine the likely underlying error signatures through a process of elimination given which $\epsilon_{\nu}^{q}$ are nonzero. While it is impossible to distinguish some combinations, e.g.~a non-QND measurement and either a measurement induced control or two-qubit error, there is sufficient information given by the mcm-rb suite to use either in a debugging cycle or to use with knowledge of the device physics to determine the likely error signatures and their underlying causes.

An important question is whether or not the mcm-rb suite can \emph{quantify} the EPM or the added EPC due to mid-circuit measurement. Since we measure the EPM by exponential fit, to quantify the error due to mid-ciruit measurement on the measured ancilla we require that the error process induce an exponential decay of ground-state probability with sequence length . While this is not generically guaranteed, we argue in the following non-QND error subsection that it applies to a wide class of error models in the small error limit.

As for quantifying the added EPC due to mid-circuit measurement, as mentioned previously from the control qubit's perspective the pair of experiments {\tt mcm-rb} and {\tt delay-rb} together form an interleaved RB (IRB) protocol. {\tt mcm-rb} interleaves a noisy identity operation on the control qubit, with the error induced by mid-circuit measurement on the ancilla. Thus, if the measurement induced error satisfies the necessary assumptions of IRB \cite{Magesan12b} we can quantify the error due to mid-circuit measurement as we would for any IRB procedure, with the error induced by measurement given by $\epsilon_{\rm IRM} = (1-\alpha_{\rm rb}/\alpha_{\rm del})/2$. We note that as with any IRB estimate of the added error, one must exercise caution with respect to quantitative accuracy as the accuracy of IRB estimates is very sensitive to the underlying errors of the reference sequence (in our case {\tt delay-rb}), as well as the nature of the interleaved error.

\section{MCM-RB in Experiment}
\label{sec:exp}

\begin{figure*}[t]
    \centering
    \includegraphics[width=\columnwidth]{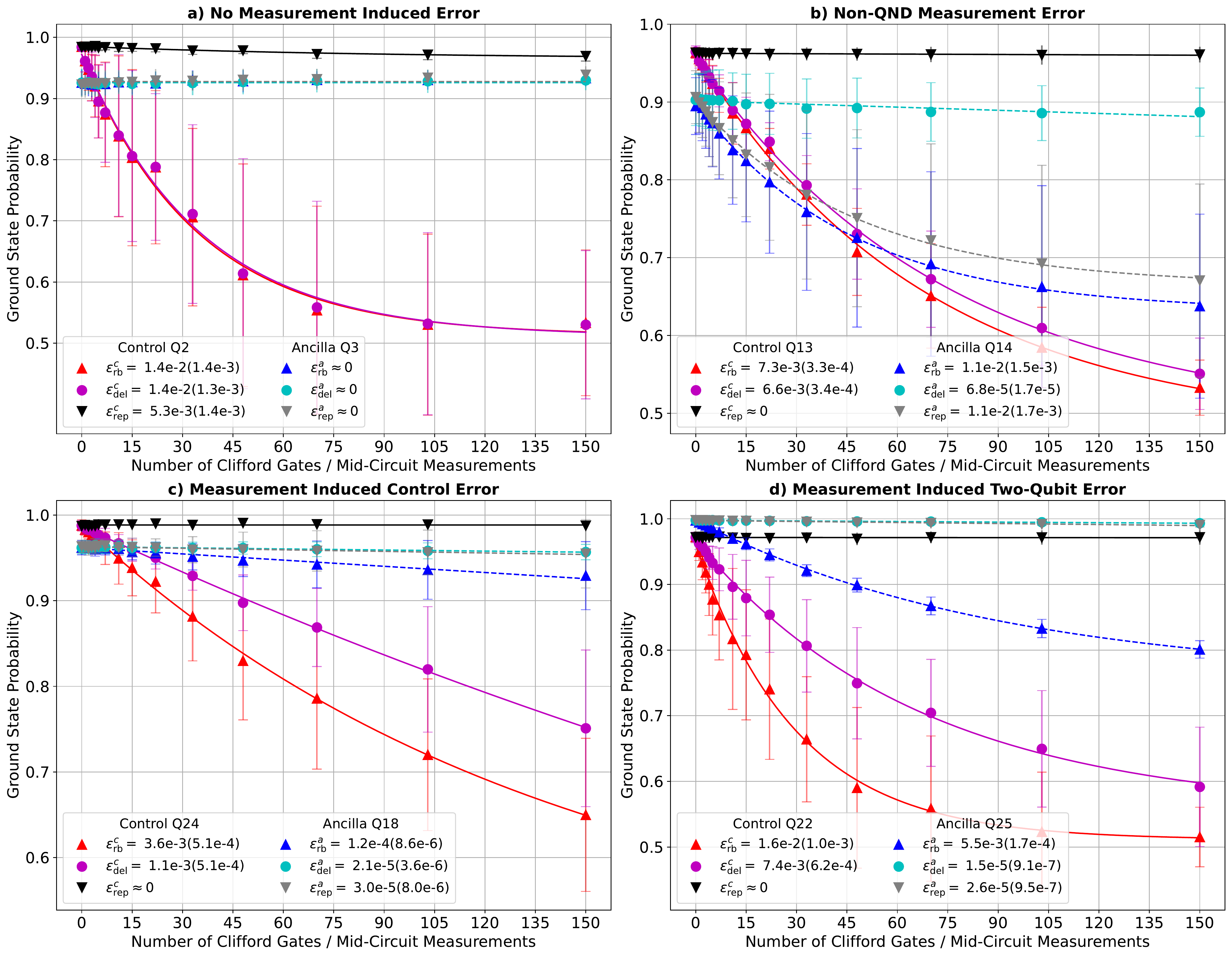}
    \caption{Error signatures of the mcm-suite on {\tt ibm\_peekskill}. For each curve, markers and error bars show the mean and one standard deviation respectively of the ground state probability over 40 random RB sequences. {\bf a) No measurement induced error} on Q2 and Q3. Even though ancilla Q3 has low readout fidelity, the mcm-rb suite shows no impact of mid-circuit measurement on either control or ancilla. {\bf b) Non-QND measurement error} for Q13 and Q14. The control Q13 is mostly unaffected by mid-circuit measurement, but the the ancilla Q14 state decays with or without Clifford RB on the control. {\bf c) Measurement induced control error} for Q24 and Q18. Decay of the control Q15 is greatly enhanced by mid-circuit measurement, but the ancilla Q12 is unaffected. {\bf d) Measurement induced 2-qubit error} for Q22 and Q25. For this, which we believe is a measurement induced collision, the decay of the control Q22 is greatly enhanced by mid-circuit measurement, and the ancilla Q25 decays only for the {\tt mcm-rb} protocol.}
    \label{fig:peekskill}
\end{figure*}

In this section, we demonstrate the practical application of the mcm-rb suite on the IBM Quantum Falcon R8 device {\tt ibm\_peekskill}. To showcase the scalability of simultaneous mcm-rb, the mcm-rb suite experiments were performed in parallel on 5 ancilla-control qubit sets for a total of 17 qubits operating simultaneously. Two distinct 17-qubit configurations on {\tt ibm\_peekskill} were considered, such that 23 of the 27 qubits on {\tt ibm\_peekskill} were studied. For further details see Appendix \ref{app:expdet}, and for complete mcm-rb suite data on all 23 qubits see Appendix \ref{app:fulldata}. Experimental data and Jupyter notebooks to reproduce the figures are available at \cite{zenodo}.

Generally, all measured ancilla qubits had weak non-QND error, with two showing considerable decay after the longest sequence (150 mid-circuit measurements). Most control qubits showed some measurement induced control error, and for many control-ancilla pairs there was evidence of measurement induced two-qubit error. As such it was not always possible to distinguish between these two error signatures, but in a few cases the distinction was significant enough to be conclusive. There were no consistent patterns observed across the device. In the following, where possible we present an example of each error signature from Table \ref{tab:esigs} using mcm-rb data taken on {\tt ibm\_peekskill}.

\subsection{No Measurement Induced Error}

The trivial error signature occurs when the EPC for {\tt mcm-rb} and {\tt delay-rb} are indistinguishable from one another, the EPC is zero for {\tt mcm-rep}, and the EPM is zero for all three experiments in the mcm-rb suite. In this case, interleaving mid-circuit measurements has no effect on either the control or ancilla qubit, which is the desired outcome for most applications. An example of no measurement induced error from {\tt ibm\_peekskill} is shown in Fig.~\ref{fig:peekskill}a). This pair of qubits was specifically chosen to also highlight that non-unity readout fidelity on the ancilla qubit does not impact the mcm-rb suite.

\subsection{non-QND Measurement}

A quantum non-demolition (QND) measurement is one for which the measurement operator commutes with the system Hamiltonian, and leaves the system in the logical eigenstate corresponding to the reported measurement outcome \cite{Braginsky80}. As such, non-QND measurement is commonly used as a catch-all term to describe any error that changes the state of the system from that reported, though this is only a subset of possible non-QND errors. One example is an error process that has a finite probability of flipping the state of the qubit after measurement. 

However, measurements that project the system onto an eigenbasis that is not the logical basis, i.e.~the logical and measurement bases are misaligned, are also non-QND. A protocol based on repeated measurements with no delays can detect non-QND errors such as measurement-induced state flips \cite{Chen21IBM}, but will be insensitive to errors due to logical and measurement basis misalignment. This insensitivity is due to the fact that from the measurement's perspective misalignment errors are not an error, and repeated measurement with no delay will repeatedly project the system into the same measurement basis state, with no probability of a state flip. Thus, there would be no decay of the ``ground-state probability''. Only the first measurement, with preparation in the logical ground state, shows any evidence of the basis misalignment, but the error induced at this sequence length is functionally indistinguishable from SPAM.

On the other hand, as {\tt mcm-rb} and {\tt mcm-rep} have delays on the ancilla between repeated measurements to accommodate control Clifford gates, there is time for the ancilla logical Hamiltonian to evolve the system out of a measurement basis state. This results in a finite probability of a measurement basis state flip after each mid-circuit measurement, and thus nonzero $\epsilon_{\rm rb}^a$ and $\epsilon_{\rm rep}^a$, such that the mcm-rb suite can detect both kinds of non-QND error. Nevertheless, in our experimental system the mid-circuit measurements have been tuned up to mitigate the impact of Stark shifts and dephasing due to residual photons in the measurement resonator \cite{Gambetta06}. This removes a major source of misalignment errors present in our system \cite{Govia15}, though others may still persist \cite{Pommerening20}.

For a generic non-QND error, as there is no unitary twirl applied to the ancilla qubit we cannot guarantee exponential decay of its ground-state probability. As a result, the EPM estimated by an exponential fit may not be a faithful quantifier of the true EPM. However, in the appropriate limit exponential decay can be obtained if the error model after each mid-circuit measurement is the same, and results in a finite probability of the qubit leaving its initial state. Such error models are common, e.g.~arising due to spurious coherent or incoherent qubit transitions driven by the measurement pulse, or the action of the logical Hamiltonian during the delay time between measurements with misaligned logical and measurement bases. While transitions out of the computational subspace, i.e.~leakage \cite{Sank16,Khezri22}, often lead to a similar exponential decay of the ancilla ground state probability, care must be taken to ensure that the qubit reset between shots is able to remove leaked population.

If the probability of a state flip after each measurement is $p$, then for an {\tt mcm-rb} and {\tt mcm-rep} sequence of length $N$ the ground state probability for the terminal measurement is the probability of an even number of state flips during the sequence, which is given by
\begin{eqnarray}
    P_{\rm GS} = \frac{(1-p)^{N+2} - p^{N+2-w}(1-p)^w}{1-2p} \label{eqn:PGS}
\end{eqnarray}
with $w = 0$ for even $N$ and $w = 1$ for odd $N$ (see Appendix \ref{app:expdet} for a derivation). In the limit of a small error such that $p \ll 1$ this becomes approximately a single exponential decay with $P_{\rm GS} \propto (1-p)^{N}$. When this is the case, an exponential fit to the data allows us to quantify the amount of induced error per measurement. 

An example of a non-QND measurement error signature from {\tt ibm\_peekskill} is shown in Fig.~\ref{fig:peekskill}b). The ancilla decay clearly shows the tell-tale signature of a non-QND error: $\epsilon_{\rm rb}^{a} \approx \epsilon_{\rm rep}^{a} > 0$, and we note that these decay curves are well fit by an exponential. While there is weak decay of the ancilla for long {\tt delay-rb} sequences, with $\epsilon_{\rm del}^{a} \neq 0$, this is three orders of magnitude less than $\epsilon_{\rm rb}^{a}$, such that it is clear the measurement is negatively impacting the ancilla. The control EPCs with and without mid-circuit measurements are not indistinguishable, indicating that there may also be some measurement induced error on the control qubit for this qubit pair.

\subsection{Measurement Induced Control Error}

The mcm-rb suite was designed to detect any errors induced by the measurement on the control qubit, and this error signature is indicative of an induced error that impacts \emph{only} the control qubit, such that the ancilla qubit can be ignored as it is unaffected. This error signature can be understood as an additional error interleaved between the non-ideal Clifford gates on the control qubit. Assuming the usual requirements of IRB hold for the measurement induced error \cite{Magesan12b}, e.g.~it is Markovian and at most weakly gate dependent, then the standard IRB procedure can be used to compare the decay of the {\tt mcm-rb} and {\tt delay-rb} experiments to quantify the error induced on the control qubit by mid-circuit measurement. Note that it is important to use {\tt delay-rb} as the reference sequence to capture the control-error due to interleaving a long delay between Clifford gates, and the magnitude of this delay error can be quantified using IRB comparing {\tt delay-rb} to a standard RB experiment with no delay.

An example of a measurement induced control error signature from {\tt ibm\_peekskill} is shown in Fig.~\ref{fig:peekskill}c). For the control qubit EPCs, $\epsilon_{\rm rb}^{c}$ is more than a factor of three larger than $\epsilon_{\rm del}^{c}$, indicating a significant impact of mid-circuit measurement on the control qubit. The EPMs for the ancilla qubit are all almost negligible, except for $\epsilon_{\rm rb}^{a}$, which indicates there may also be evidence for a weak two-qubit measurement induced error or RB cross-talk error.  

While the physical origin of this error cannot be determined by the mcm-rb suite, given the nature of the hardware platform and the fact that $\epsilon^c_{\rm rep} \approx 0$, it is likely due to either a measurement induced Stark shift, or weak cross-dephasing on the control qubit. The power of the mcm-rb suite is that it can quickly identify all such issues across an entire chip, which can then be explored in more detail with slower techniques to determine their origin. 

\begin{figure}[t!]
    \centering
    \includegraphics[width=0.75\columnwidth]{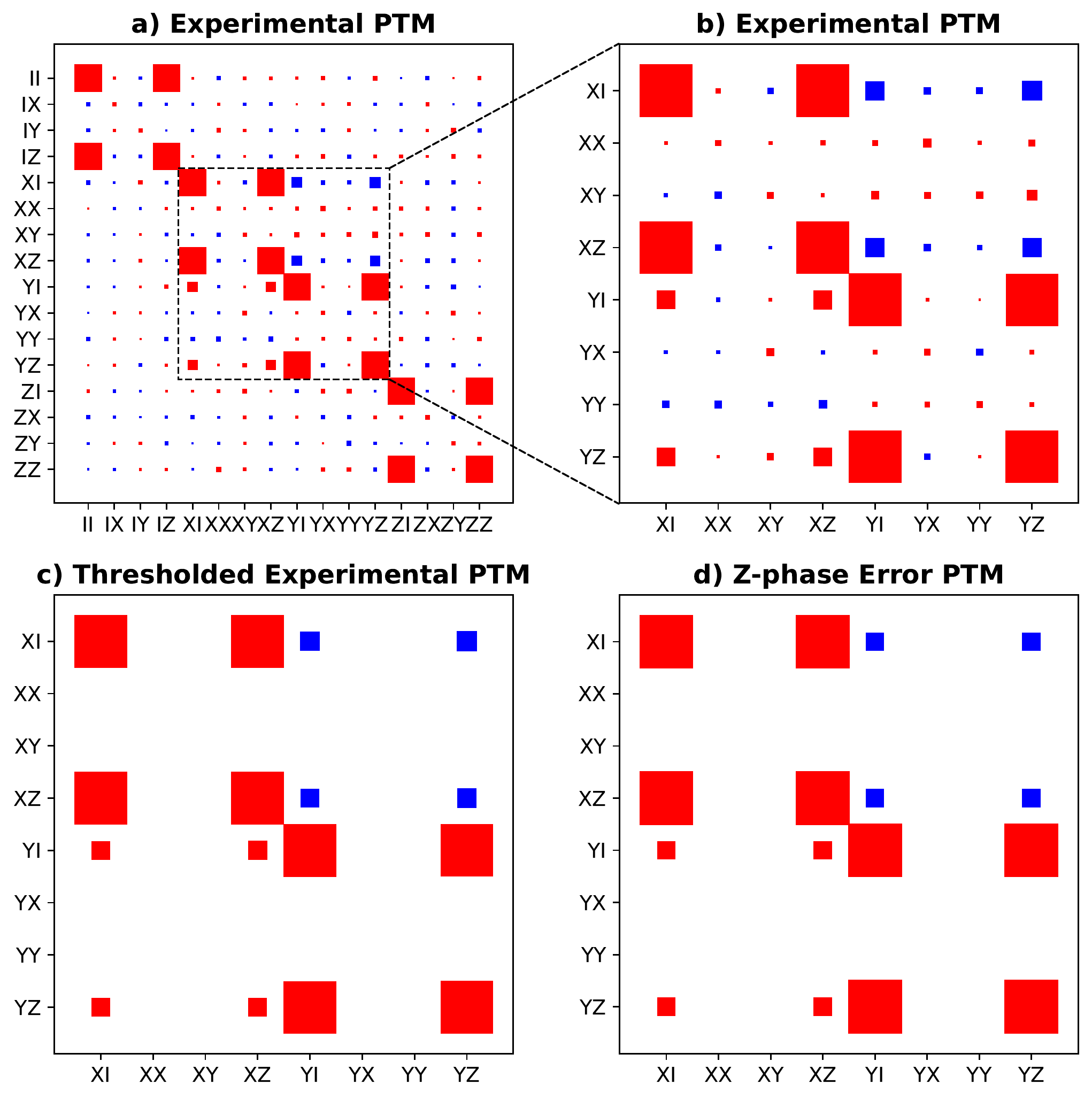}
    \caption{Results of mid-circuit measurement tomography applied to control Q24 and ancilla Q18. Panel a) shows the Pauli transfer matrix for the two-qubit channel (on Q24 and Q18) during measurement of Q18, and panel b) is a zoom in of the interior blocks. Panel c) is a thresholded version of panel b), where only PTM elements with magnitude greater than $\sqrt{6\epsilon_{\rm IRB}}$ are plotted. Panel d) presents the PTM for a simulated ideal measurement on ancilla Q18 and a Z-phase error on control Q24 with average gate infidelity of $\epsilon_{\rm IRB}$.}
    \label{fig:tomo}
\end{figure}

In the case of this qubit pair, we performed a follow-up experiment that interleaved data collection for the mcm-rb suite protocols and a mid-circuit measurement tomography protocol. From the results of the mcm-rb suite we obtained an estimate of the control error induced by measurement of $\epsilon_{\rm IRB} = 1.7{e-}3 \pm 1.0{e-}3$, which unfortunately is a good example of the large uncertainty bounds on IRB estimates. From the results of the measurement tomography we obtained the Pauli transfer matrix (PTM) shown in Fig.~\ref{fig:tomo}a), which shows the expected signal for the ideal channel (the dominant block diagonal structure in red) along with many spurious non-zero elements arising from the error channel.

 Zooming into the interior blocks in Fig.~\ref{fig:tomo}b), the largest error source has a structure indicative of coherent Z-phase error induced by Stark shift. For a Z-phase error of angle $\theta$ induced on the control qubit by ancilla measurement, we would have that $\epsilon_{\rm IRB} = (1-4\cos(2\theta))/3$. For the PTM the non-zero elements due to the Z-phase error all have the same magnitude, given by
\begin{eqnarray}
    \left|R_{\iden\hat{Y},\iden\hat{X}}\right| = \left|\sin(2\theta)\right| \approx \sqrt{6\epsilon_{\rm IRB}},
\end{eqnarray}
where in the last expression we have used the second order in $\theta$ series expansions of $\epsilon_{\rm IRB}$ and $R_{\iden\hat{Y},\iden\hat{X}}$ to relate the two quantities.

Only plotting PTM elements with magnitude larger than $\sqrt{6\epsilon_{\rm IRB}}$ (using the mean of the estimate for $\epsilon_{\rm IRB}$) produces Fig.~\ref{fig:tomo}c). This closely matches the PTM shown in Fig.~\ref{fig:tomo}d) for a simulated Z-phase error with a $\theta$ calculated from $\epsilon_{\rm IRB}$. This demonstrates how the quantitative error benchmarking of the mcm-rb protocol can be used to help interpret the results of the more detailed mid-circuit measurement tomography, and in so doing obtain both the error magnitude and its nature.

\subsection{Measurement Induced Two-Qubit Error}

Unlike the error signatures we have thus far considered, there are sufficiently diverse measurement induced two-qubit errors that they will not all result in the same error signature. We do expect that any measurement induced two-qubit error will result in a faster decay of the control qubit for {\tt mcm-rb} compared to {\tt delay-rb}. Unfortunately, the control {\tt mcm-rb} decay is not guaranteed to be exponential as only the control-qubit is twirled. However, from the simultaneous RB protocol \cite{Gambetta12} we know that in the limit of small two-qubit error the decay of a single-subsystem Clifford twirl will still be approximately exponential, such that we can again quantify the added error on the control using an IRB procedure comparing {\tt mcm-rb} and {\tt delay-rb}.

The signature of the ancilla decay curves is not consistent across the various error models that fall under measurement-induced two-qubit error. For example, if the error is a coherent excitation exchange between control and ancilla, then we would expect to see decay of the ancilla ground state probability only for the {\tt mcm-rb} protocol but not the {\tt mcm-rep} protocol, since the ancilla and control are both initialized in the ground state. On the other hand, a double excitation error (i.e.~ an $XX$-gate) induced by measurement would result in finite ancilla EPM for both {\tt mcm-rb} and {\tt mcm-rep}, while a correlated phase error (i.e.~ a $ZZ$-gate) would not impact the ancilla state.

An example of a measurement induced two-qubit error signature from {\tt ibm\_peekskill} is shown in Fig.~\ref{fig:peekskill}d). In this case, it is clear that this is a two-qubit error as we have that both a substantial $\epsilon^{a}_{\rm rb} > 0$ and $\epsilon^{c}_{\rm rb} > \epsilon^{c}_{\rm del}$, with all three decay curves well fit by exponential functions. For this particular two-qubit error, $\epsilon^{a/c}_{\rm rep}$ is negligible compared to $\epsilon^{a}_{\rm rb}$. As we explain in more detail with a numerical simulation model in Section \ref{sec:sims}, we attribute this particular kind of two-qubit error signature to a measurement induced collision. The ancilla qubit is Stark-shifted by the photons in the measurement cavity to a frequency close enough to the control qubit such that near-resonant excitation exchange can occur. The tell-tale signature that this is likely a collision is the fact that $\epsilon^{a}_{\rm rb} \gg \epsilon^{a}_{\rm rep}$, indicating that for a control qubit in its ground state the ancilla is not impacted.

\subsection{RB Cross-talk}

Finally, we briefly discuss an error signature detectable by the mcm-rb suite, but which has nothing to do with mid-circuit measurement. If the implementation of the Clifford gates on the control qubit impacts the ancilla qubit, i.e.~if there is cross-talk between the qubits for single qubit gates, then the {\tt mcm-rb} and {\tt delay-rb} curves for the ancilla will likely decay. However, due to the lack of a Clifford twirl on the ancilla, the mcm-rb suite cannot say much quantitatively about this error. By design the mcm-rb suite is meant to benchmark errors induced by measurement on the control qubit(s), and other protocols exist to benchmark \cite{Gambetta12,McKay20} or characterize \cite{Rudinger21} cross-talk. 

\section{Numerical Simulation of MCM-RB}
\label{sec:sims}

In the following subsections, for each non-trivial error signature of Table \ref{tab:esigs} we use numerical simulation to study example physical error mechanisms that lead to the error signature. Our simulations are performed using the quantum circuit simulator in Qiskit Aer \cite{Qiskit}, which natively supports error processes such as the depolarizing channel and decoherence generated by qubit $T_1$ decay and $T_2$ dephasing. Additionally, one can add custom error processes either as unitary gates or via their Kraus decomposition, and we make use of this functionality for several of the measurement induced errors considered in the subsequent subsections.

In addition to the measurement induced error, our simulations add depolarizing error to each single qubit gate on the control qubit, as well as an amplitude and phase damping channel to the control qubit for each ancilla measurement. The latter is equivalent to the decoherence generated by control qubit $T_1$ and $T_2$ decoherence during the measurement, with parameters chosen to be representative of our experimental setup. For further details see Appendix \ref{app:simdet}.

\subsection{Non-QND Measurement}

\begin{figure}[t]
    \centering
    \includegraphics[width=0.5\columnwidth]{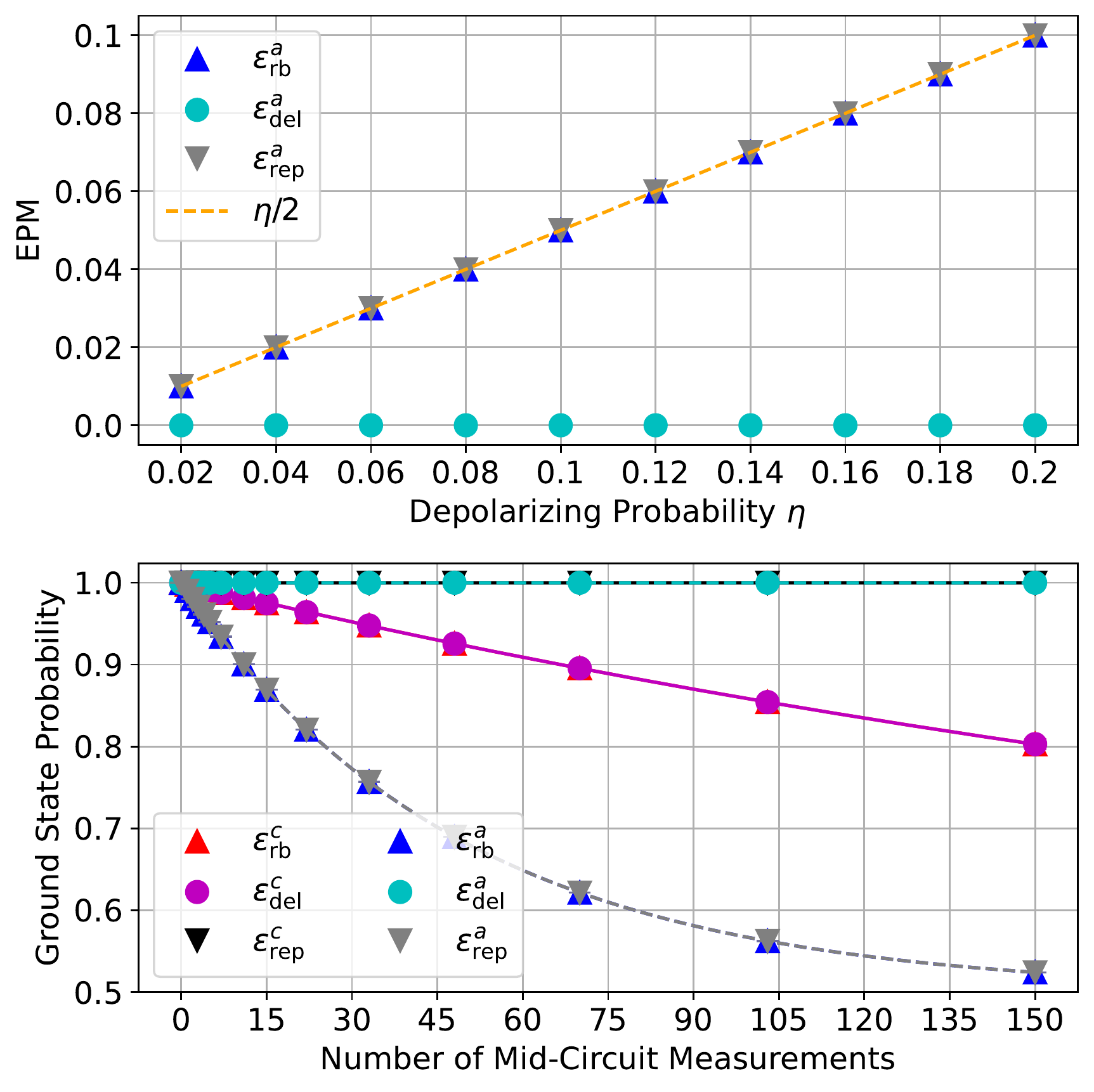}
    \caption{{\bf Non-QND Measurement Simulations.} (Upper panel) EPM for the ancilla qubit for the three mcm-rb suite experiments. The EPM for {\tt mcm-rb} and {\tt mcm-rep} accurately estimates the simulated EPM of $\eta/2$, shown in the dashed orange line. (Lower panel) Example mcm-rb suite decay curves for a non-QND measurement error due to a depolarizing channel with depolarizing probability $\eta = 2\%$. In addition to the ancilla depolarizing channel, we apply an amplitude and phase damping channel corresponding to dissipation with $T_1 = 345~\mu$s, $T_2 = 280~\mu$s, and duration $t_m = 0.71~\mu$s on the control qubit.}
    \label{fig:sims_nonqnd}
\end{figure}

To model non-QND measurement, we take a simple approach and consider the application of a depolarizing channel after each mid-circuit measurement. A single-qubit depolarizing channel acts as
\begin{eqnarray}
    \mathcal{E}_{\rm dep}(\rho) = (1-\eta)\rho + \eta\frac{\iden}{2}, \label{eqn:dep}
\end{eqnarray}
with $\eta$ the depolarizing probability. For our simulations we scan $\eta$ from $2\%$ to $20\%$, as shown in Fig.~\ref{fig:sims_nonqnd}. The lower panel shows an example of the expected mcm-rb suite decay curves for a non-QND measurement error, in this case for $\eta = 2\%$. The upper panel shows that the EPM for both {\tt mcm-rb} and {\tt mcm-rep} accurately estimate the simulated error per measurement, which for the depolarizing channel is $\eta/2$.

\subsection{Measurement Induced Control Error}

\begin{figure*}[t!]
    \centering
    \includegraphics[width=\columnwidth]{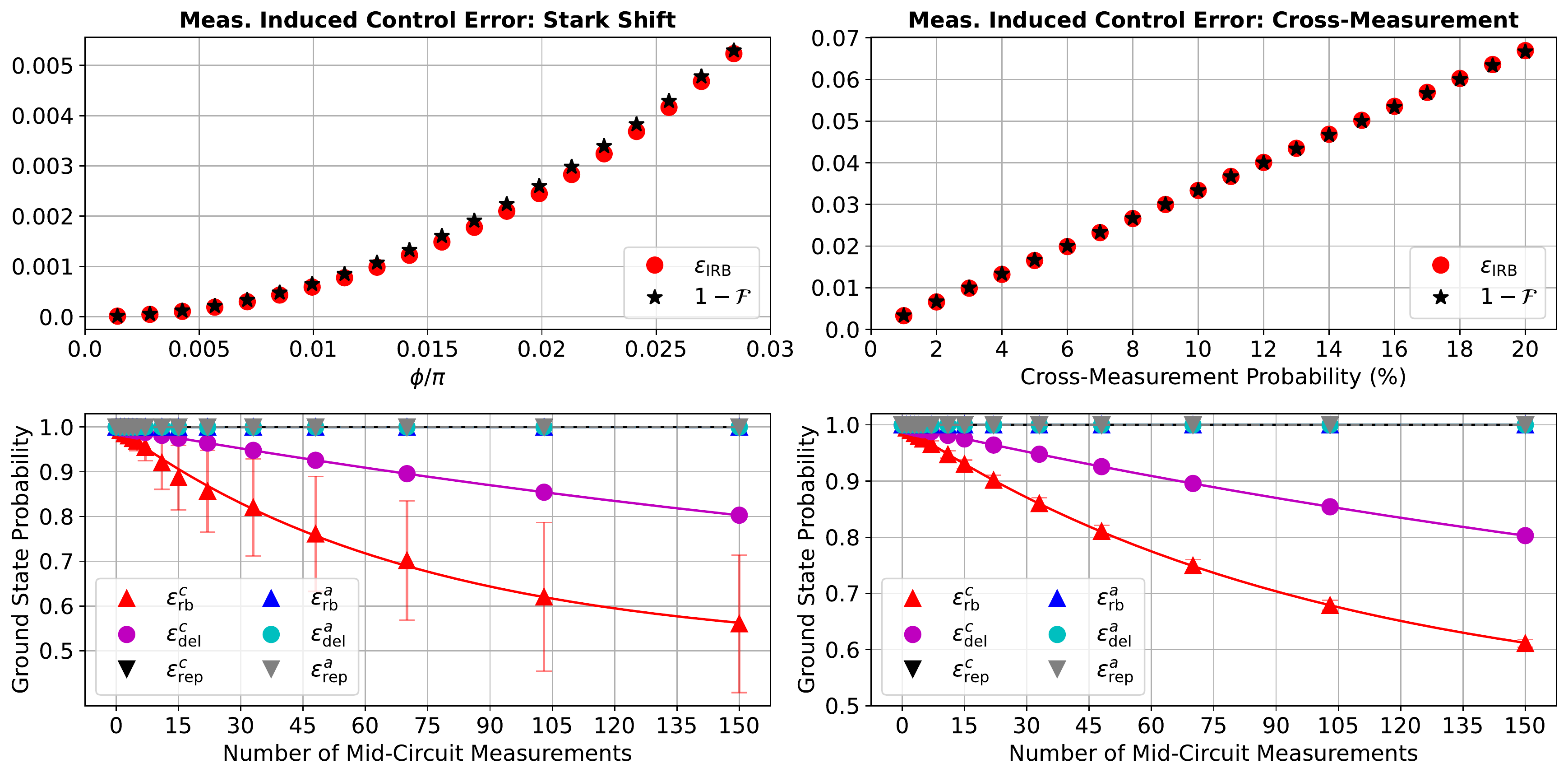}
    \caption{{\bf Measurement Induced Control Error.} Left column: $Z$-phase error due to control-qubit Stark shift. Right column: Error due to a cross-measurement on the control qubit as described by the channel of Eq.~(\ref{eqn:cm}). For both simulations in addition to the aforementioned errors we apply an amplitude and phase damping channel corresponding to dissipation with $T_1 = 345~\mu$s, $T_2 = 280~\mu$s, and duration $t_m = 0.71~\mu$s on the control qubit. The upper panels shows the added error due to measurement as estimated by IRB, $\epsilon_{\rm IRB} = (1-\alpha_{\rm rb}/\alpha_{\rm del})/2$, compared to the average gate infidelity $1-\mathcal{F}$ of the simulated process. In both cases there is good agreement between the estimated and exact values. The lower panels show example mcm-rb suite decay curves, for $\phi/\pi \approx 0.03$ and $p_m =0.01$. Despite their similar decay rates, note the considerable spread in the control-qubit {\tt mcm-rb} decay curve for the coherent error compared to the incoherent error. }
    \label{fig:sim_control}
\end{figure*}

We consider two models for physical error mechanisms that induce control qubit errors. The first is a measurement-induced Stark shift, which adds a $Z$-phase error to the control qubit after every measurement via the unitary $\hat{U}_{\rm Stark} = e^{-i\phi\hat{\sigma}_z}$. For a cQED system such as {\tt ibm\_peekskill}, this can occur when readout photons intended for the ancilla-qubit resonator populate the control-qubit resonator. Then, via the dispersive interaction $\hat{H} = \chi \hat{\sigma}^c_z\hat{n}$ between the control-qubit and its readout resonator, the control frequency is Stark shifted by $2\chi\bar{n}$ during measurement, where $\bar{n}$ is the average number of photons in the resonator. For a mid-circuit measurement of duration $t_m$ this leads to a Stark-phase error with $\phi = 2\chi\bar{n}t_m$.

The second model is cross-measurement, where with some probability $p_m$ the measurement of the ancilla also strongly measures the control qubit. This error is described by the quantum channel, $\mathcal{E}^c_{p_m}$, that completely dephases the control qubit with probability $p_m$. The Kraus representation of this channel is given by
\begin{eqnarray}
     &\hat{K}_0 = \sqrt{p_m}\ketbra{0},~\hat{K}_1 = \sqrt{p_m}\ketbra{1},~\hat{K}_2 = \sqrt{1-p_m}\hat{\mathbb{I}}. \label{eqn:cm}
\end{eqnarray}
In a continuous time model this error could also be described by dephasing on the control qubit during the measurement with a rate $\gamma$ defined by $e^{-\gamma t_m} = 1-p_m$. In a cQED system, this error can occur due to readout photons that leak into the control-qubit resonator, and is also possible in an ion trap system due to scattering of the readout laser pulse or photons fluoresced by the ancilla-qubit \cite{Gaebler21}.

The results of our simulations are shown in Fig.~\ref{fig:sim_control}, with the left column showing the results for the Stark shift error model and the right column for the cross-measurement error model. The lower panels show examples of the mcm-rb suite decay curves 
for these error models, and it is important to highlight that despite very different underlying physics, they produce the same error signature: $\epsilon^{c}_{\rm rb} \gg \epsilon^{c}_{\rm del}$ with all other EPC/M zero. This is to be expected as they are both errors that impact only the control qubit.

The upper panels show the error induced on the control qubit per mid-circuit measurement as estimated by IRB ($\epsilon_{\rm IRB}$), and compare to the average gate infidelity, i.e.~$1-\mathcal{F}$, with $\mathcal{F}$ the average gate fidelity which can be calculated exactly for these error channels. In both cases the IRB estimate is reasonably accurate, but it is noticeably less so for the coherent Stark shift error model. This highlights the caution necessary when using IRB, especially for coherent error and if a high degree of accuracy is required. Comparing the mcm-rb suite decay curves of the lower panels, it is unsurprising that the IRB estimate is less accurate for the coherent Stark-shift error, given the much larger spread in the control-qubit decay curves for {\tt mcm-rb} observed for that error model.

\subsection{Measurement Induced Two-Qubit Error}
\begin{figure}[t!]
    \centering
    \includegraphics[width=0.5\columnwidth]{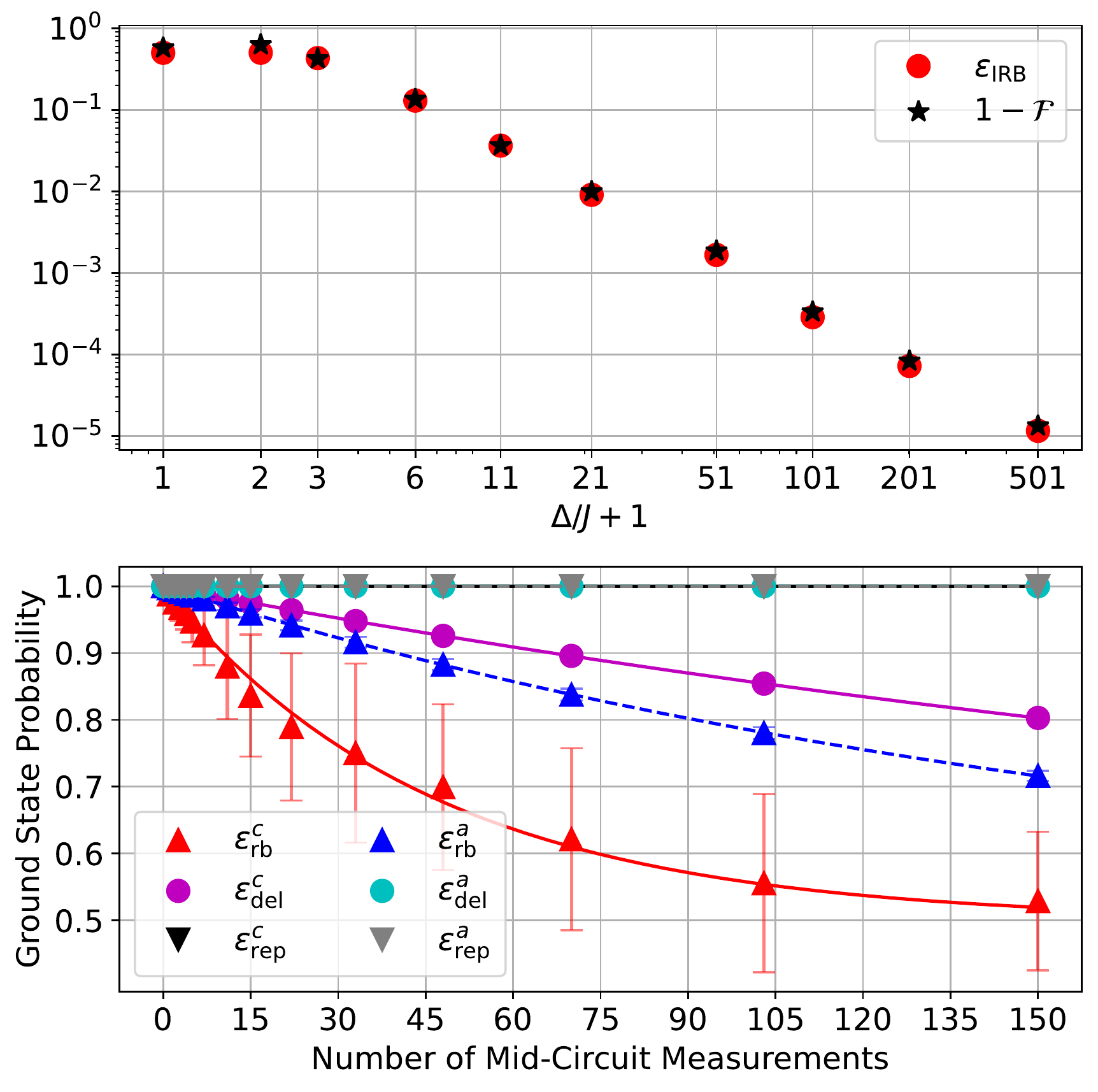}
    \caption{{\bf Measurement Induced Two-Qubit Error.} The measurement induced collision error model of Eq.~(\ref{eqn:col}) parameterized by the ratio of the detuning and coupling, $\Delta/J$. We also apply an amplitude and phase damping channel corresponding to dissipation with $T_1 = 345~\mu$s, $T_2 = 280~\mu$s, and duration $t_m = 0.71~\mu$s on the control qubit. (Upper panel) Added error due to measurement as estimated by IRB, $\epsilon_{\rm IRB}$, compared to the average gate infidelity $1-\mathcal{F}$. Note that the values of the $x$-axis are shifted by one to accommodate the log-log scale. (Lower panel) Example mcm-rb suite decay curves for a measurement induced collision with $\Delta/J = 20$.}
    \label{fig:sim_col}
\end{figure}

From the broad class of possible two-qubit errors we consider one that is physically motivated by the superconducting hardware platform, and has an interesting error signature. In particular, we explore the impact of a measurement induced collision, where the measurement induced Stark shift on the ancilla qubit brings it close to resonance with the control qubit. We consider a minimal model for this system that is platform agnostic, described by a coupled control-ancilla system. The Hamiltonian describing their interaction is
\begin{eqnarray}
    \hat{H} = \frac{\Delta}{2}\hat{\sigma}_{z}^a + J\left(\hat{\sigma}^a_-\hat{\sigma}^c_+ + \hat{\sigma}^a_+\hat{\sigma}^c_-\right), \label{eqn:col}
\end{eqnarray}
where the qubits are in a frame rotating at the frequency of the control qubit such that $\Delta = \omega_a - \omega_c$, with $\omega_a$ the Stark-shifted frequency of the ancilla qubit. To implement our error model in simulation, after every measurement we add the unitary error $\hat{U}_{\rm Col} = e^{-i\hat{H}t_m}$, which corresponds to evolving the system for a time $t_m$ under the evolution of the collision Hamiltonian. 

Figure \ref{fig:sim_col} shows the results of these simulations. As in our other simulations, the upper panel shows that the error added by measurement can be accurately estimated using an IRB procedure to compare $\epsilon^{c}_{\rm rb}$ and $\epsilon^{c}_{\rm del}$. See Appendix \ref{app:simdet} for further information on our calculation of the average gate infidelity, $1-\mathcal{F}$, of the effective single qubit channel on the control qubit induced by this two-qubit error channel. We note that even though the IRB prediction is accurate, for $\Delta/J < 5$ the decay curves are not well fit by exponential functions, as the error induced by the two-qubit channel is too large for the single-subsystem twirl that is performed \cite{Gambetta12}.

The lower panel shows an example of the expected mcm-rb suite decay curves ($\Delta/J = 20$), with $\epsilon^{c}_{\rm rb} \gg \epsilon^{c}_{\rm del}$ and finite $\epsilon^{a}_{\rm rb}$, all of which display exponential decay. It is the finite $\epsilon^{a}_{\rm rb}$ that distinguishes a two-qubit error from an error only on the control qubit. As it requires one qubit to be at least partially excited, $\epsilon^{a}_{\rm rep} = 0$ for a measurement induced collision, but other two-qubit error sources may have finite $\epsilon^{a}_{\rm rep}$. 

\section{Discussion}
\label{sec:Disc}

For our demonstrations of the mcm-rb suite we have exclusively focused on the scenario where the ancilla qubit is initially prepared in the ground state. An equivalent set of experiments could be performed with the ancilla qubit prepared in the excited state, and this would return different results if the effective error channel on the control qubit depends on the ancilla state. On its own this is not problematic, and one could simply repeat the mcm-rb suite for both ancilla initial states, or randomize if the average channel is of more interest. However, care must be taken if measurement can induce ancilla state flips, as the change in the instantaneous control-qubit error channel due to the change in ancilla state would result in an overall non-Markovian control-qubit error channel.

As an example, consider the experimentally relevant situation of an ancilla with relaxation characterized by a timescale $T_1$. If the ancilla is initialized in the excited state, then for the initial duration of an RB sequence ($t \ll T_1$) the effective error channel on the control qubit is approximately static, given by the error channel conditioned on the ancilla in the excited state. For sequences with duration longer than $T_1$, near the end of the sequence ($t\gg T_1$) the effective control-qubit error is again approximately static, but now given by the error channel conditioned on the ancilla in the ground state \footnote{Note that while the celebrated quantum Zeno effect plays an important role in determining exactly how long a sequence needs to be before the ancilla has likely decayed, only for measurements repeated infinitely quickly would the ancilla by pinned in its excited state in perpetuity.}.

Crucially, at some point during the sequence the control-qubit error channel changes, and thus across the total sequence the control-qubit error cannot be consistently defined by one single-qubit quantum channel. The control-qubit error is temporally correlated across the sequence in a non-trivial way, with quasi-static error that exhibits at most one switch during a given sequence. The impact of such temporal correlations on RB has been previously studied \cite{Ball16}, and in practice it does not preclude EPC estimation, but may make the estimates less reliable and require more random sequences for convergence.

In our experimental system, the $T_1$ time is a factor of 2 or 3 longer than the longest sequences we use in our mcm-rb experiments. Our system exhibits ancilla-state-dependent control-qubit error due to the presence of weak $ZZ$-coupling between many of the qubits on the device. Due to the fact that gates are calibrated with all spectator qubits in the ground state, this results in a coherent error on the control qubit only for an excited ancilla, $\hat{U}_{ZZ} = e^{-i\hat{H}_{ZZ} t_m}$, described by the Hamiltonian
\begin{eqnarray}
    \hat{H}_{ZZ} = \nu \ketbra{e}_a \otimes \hat{\sigma}_z^c, \label{eqn:ZZ}
\end{eqnarray}
where $\nu$ is the $ZZ$-coupling rate.

\begin{figure}[t!]
    \centering
    \includegraphics[width=0.5\columnwidth]{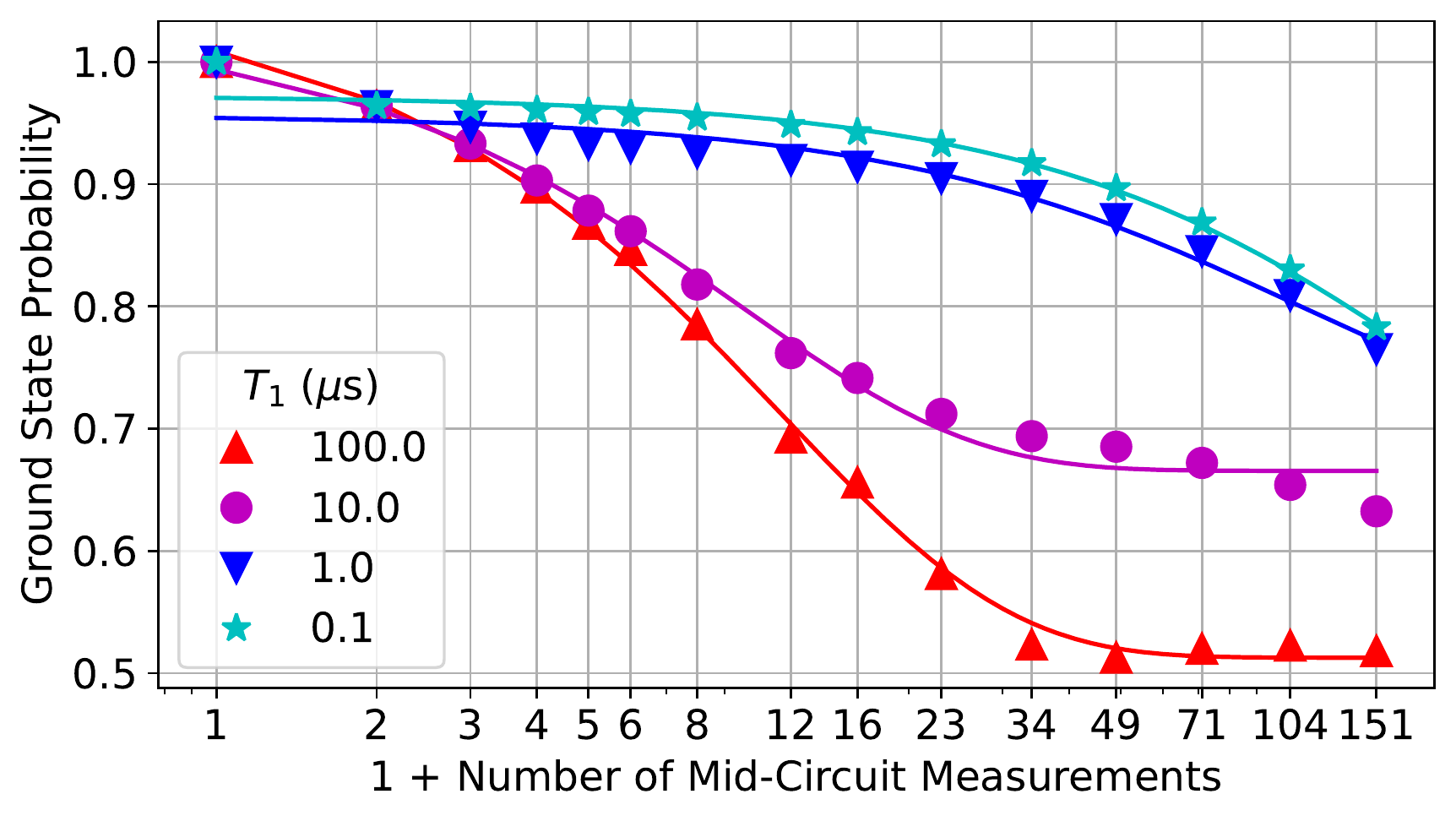}
    \caption{{\bf Non-Markovian Control Error Induced by the Ancilla.} Control-qubit decay of the {\tt mcm-rb} protocol, i.e.~$\epsilon_{\rm rb}^c$, for the $ZZ$-coupling error model of Eq.~(\ref{eqn:ZZ}) combined with ancilla relaxation decay for varying values of the relaxation timescale $T_1$. For all curves, the ancilla-qubit was initialized in the excited state, $\nu = 50$ kHz, and $t_m$ is as before. Note that the x-axis is shifted by $1$ to accommodate a value of $0$ on the log-scale.}
    \label{fig:non-markov}
\end{figure}

We simulate the impact of this error channel combined with relaxation of the ancilla-qubit on the {\tt mcm-rb} protocol with the ancilla initialized in the excited state, and the results of these simulations are shown in Fig.~\ref{fig:non-markov}. When $T_1$ is very short (e.g.~$0.1$ or $1.0~\mu$s) the ancilla relaxes almost immediately, such that only the first gates in a given sequence experience the error $\hat{U}_{ZZ}$. All but the first few sequence lengths are well fit by an exponential with an $\epsilon_{\rm rb}^c$ calculated from the control-qubit error model for the ancilla in the ground state. For very long $T_1$ (e.g.~$100~\mu$s), only the longest sequences are likely to experience ancilla relaxation. Aside from small deviations at the end, the full decay curve is well fit by an exponential with $\epsilon_{\rm rb}^c$ calculated from $\hat{U}_{ZZ}$. For intermediate $T_1$ (e.g.~$10~\mu$s) the fit quality decreases significantly for the longer sequences where an ancilla relaxation, and thus an inconsistency in the control-qubit error, is likely to occur.

As these simulations show, the non-Markovian characteristic of the control-qubit error induced by the combination of $ZZ$-coupling and ancilla relaxation reduce the reliability of the EPCs obtained from RB fitting. Though our ancilla qubits have an average $T_1 > 100~\mu$s, to avoid the complications of temporally correlated error in RB we have focused on benchmarking mid-circuit measurement with the ancilla in the ground state.

One possible way to overcome this issue would be to randomize either the initial ancilla state preparation, or randomize state re-initialization after each mid-circuit measurement. This can be done by randomly inserting identity or $\hat{X}$-gates at the start of the circuit, or after each mid-circuit measurement. A further extension would be to consider the full Pauli-twirl of the mid-circuit measurement, so that the action on the control qubits was guaranteed to be a stochastic channel \cite{Beale23}. In aggregate, the control qubit will then experience the average error channel induced by ancilla measurement, unconditioned on the ancilla state.

Similarly, we could randomize the initialization of the control qubit, which should not impact {\tt mcm-rb} or {\tt delay-rb}, but could potentially change the result of {\tt mcm-rep}. This would be the case, for example, if the measurement induced an amplitude damping error on the control qubit. We leave exploration of these extensions of the mcm-rb suite, and how their measured EPCs connect to experimentally relevant quantities, for future study.

\section{Conclusion}
\label{sec:Conc}

In this work we have presented a randomized benchmarking suite for mid-circuit measurements, whose central protocol interleaves mid-circuit measurements on an ancilla qubit between Clifford gates on a control qubit. The remaining two protocols of the suite replace either the mid-circuit measurement or the Clifford gates with idle delays of equal duration, and serve as reference experiments to enable error quantization through an interleaved randomized benchmarking procedure. As we have demonstrated on an IBM Quantum Falcon device, our benchmarking suite can be trivially extended to an entire multi-qubit chip, benchmarking multiple control and ancilla qubits simultaneously.

The mcm-rb suite classifies errors based on their error signature, which is the relationship between the RB-decay curves from the three protocols in the suite for both the control and ancilla qubits. We discussed the three major error signatures: non-QND measurement error, control-qubit error, two-qubit error; and highlighted examples of these error signatures from our deployment of the mcm-rb suite on an IBM Quantum Falcon device. Each error signature can be the result of many different physical error models, and we explored several in numerical simulation. By comparing to the average infidelity of our simulation models, we demonstrated that the mcm-rb suite can function as an IRB procedure and quantify the error added by mid-circuit measurement.

Our benchmarking suite can be readily adapted to other quantum-classical operations beyond mid-circuit measurement. These include a larger part of or even the full circuit for a stabilizer check, and real-time operations such as measurement and feed-forward \cite{Riste20}. While we have motivated mid-circuit measurement by its necessity in fault-tolerant quantum computing, many proposed near-term algorithms would benefit from this capability or the real-time operations it enables \cite{Urbanek20,Corcoles21,Botelho22,Piveteau22}. Thus, we expect the mcm-rb suite and developments upon it to also have immediate impact in characterizing devices for near-term applications.

\section*{Acknowledgements}
The device was designed and fabricated internally at IBM. We acknowledge the use of IBM Quantum services for this work, and these results were enabled by the work of the IBM Quantum software and hardware teams. Access to devices was supported by IARPA under LogiQ (contract W911NF-16-1-0114). The research of this manuscript was sponsored by the Army Research Office and was accomplished under Grant Number W911NF-21-1-0002. The views and conclusions contained in this document are those of the authors and should not be interpreted as representing the official policies, either expressed or implied, of the Army Research Office or the U.S.~Government. The U.S.~Government is authorized to reproduce and distribute reprints for Government purposes notwithstanding any copyright notation herein.

\appendix

\section{Experimental details}
\label{app:expdet}

\begin{figure}[t]
    \centering
    \includegraphics[width=0.75\columnwidth]{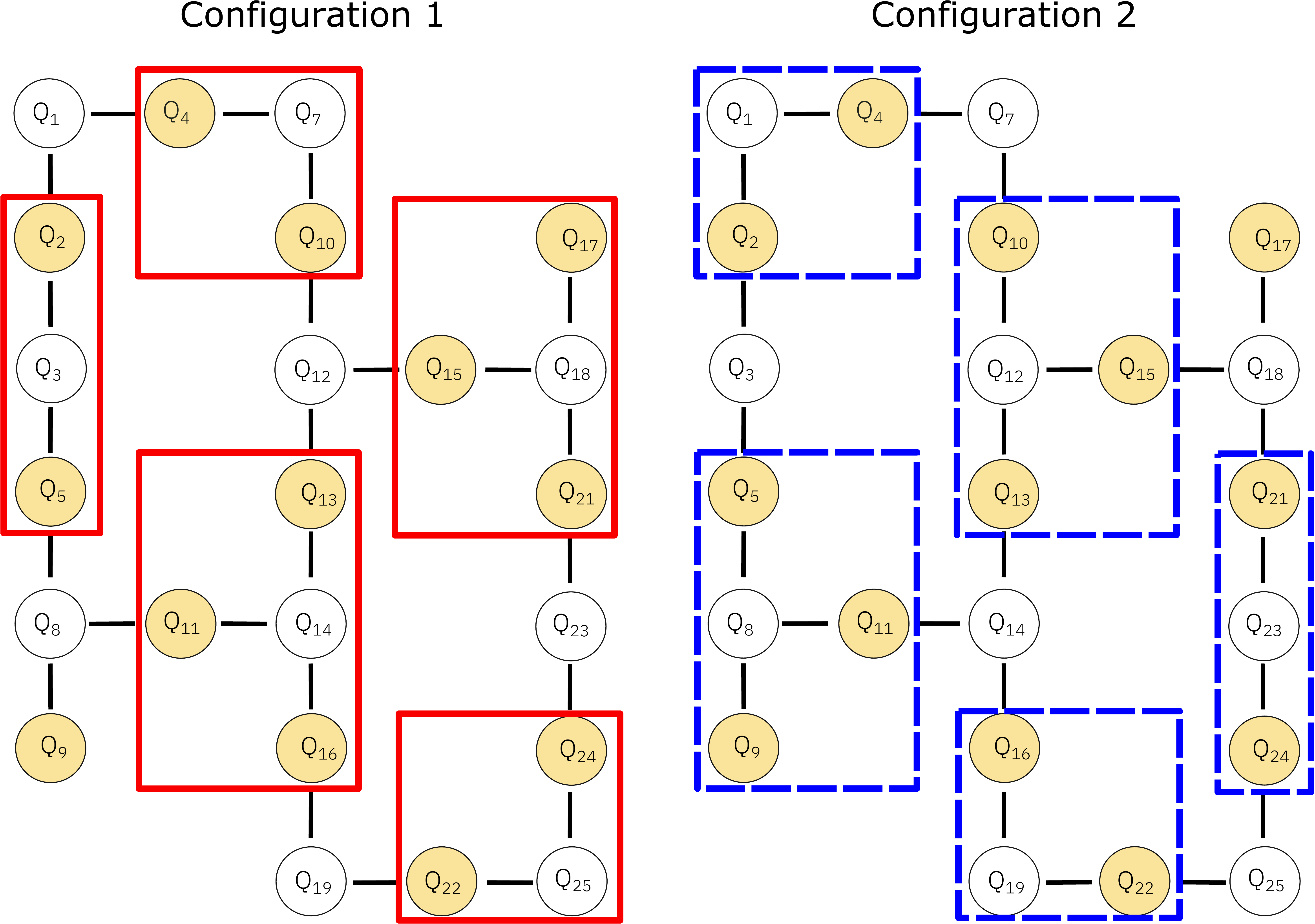}
    \caption{Configurations for the two simultaneous mcm-rb suite experiments that each involved 17 qubits from {\tt ibm\_peekskill}. Ancilla qubits are shown in white, and control qubits in yellow. Red/blue squares encompass each ancilla-controls group.}
    \label{fig:config}
\end{figure}

The mode of operation we employ to perform the mcm-rb sutie simultaneously across {\tt ibm\_peekskill} is to parallelize both over control and ancilla qubits. We break the full set of qubits into groups that each consist of one ancilla qubit and two or three control qubits. As shown in Fig.~\ref{fig:config}, with just two configurations of 17 qubits each we can cover 23 of the 27 qubits on {\tt ibm\_peekskill}. For each protocol that involved Clifford RB, we used 40 random sequences for each of the 15 sequence lengths, and took 1024 shots for each sequence. Each configuration therefore consisted of 1800 circuits total, which was broken into 5 jobs that were run sequentially.

In section \ref{sec:exp} B we consider a model for non-QND measurement error where after each mid-circuit measurement the state of an ancilla qubit has probability $p$ of flipping to the orthogonal state. As we start in the ground-state the probability we end in the ground state is the probability that there have been an even number of state flips in the sequence. This probability is the sum of the probabilities of all possible even numbers of flips, which for $N$ mid-circuit measurements is given by
\begin{eqnarray}
    P_{\rm GS} = \sum_{k}^{\lfloor N/2\rfloor} p^{2k}(1-p)^{N-2k}.
\end{eqnarray}
The closed form expressions for $N$ even and $N$ odd are
\begin{eqnarray}
    P_{\rm GS} = \left\{\begin{array}{cc}
    \frac{(1-p)^{N+2} - p^{N+2}}{1-2p} & N~{\rm even}\\ 
    \frac{(1-p)\left((1-p)^{N+1} - p^{N+1}\right)}{1-2p} & N~{\rm odd} \end{array}\right. ,
\end{eqnarray}
which can be unified into the single expression given in Eq.~(\ref{eqn:PGS}).

\section{Numerical simulation details}
\label{app:simdet}

Jupyter notebooks implementing our simulations can be found at \cite{zenodo}. The Qiskit Aer simulator we employ is a gate-based simulator of a quantum circuit, with non-unitary noise channels implemented in Kraus form. To simulate the action of mid-circuit measurement, we apply a completely dephasing channel to the ancilla qubit, $\mathcal{E}_{\rm m}^a$, which is described by the Kraus operators $K_0 = \ketbra{0}$ and $K_1 = \ketbra{1}$. As we discard the outcomes of our mid-circuit measurements in experiment, $\mathcal{E}_{\rm m}^a$ is equivalent to the action of ideal mid-circuit measurement in the ensemble average.

To add errors to the mid-circuit measurement, we sandwich each application of $\mathcal{E}_{\rm m}^a$ with a pre-measurement and post-measurement two-qubit noise channel, which we label $\mathcal{E}_{\rm pre}$ and $\mathcal{E}_{\rm post}$. Table \ref{tab:sims_models} shows the choice of $\mathcal{E}_{\rm pre}$ and $\mathcal{E}_{\rm post}$ for each of our simulations. In addition to the errors before and after measurement, all single-qubit gates in our simulation have a depolarizing error channel ($\eta = 10^{-3}$) applied after the action of the gate.

\begin{table}[h]
    \centering
    \begin{tabular}{|l|l|l|}
    \hline
         {\bf Simulation} &  $\mathcal{E}_{\rm pre}$ &  $\mathcal{E}_{\rm post}$ \\
         \hline 
         \hline
         non-QND measurement & $\mathcal{E}^c_{T_1, T_2}$ & $\mathcal{E}^a_{\rm dep}$ \\
         \hline
         Stark shift & $\hat{U}^c_{\rm Stark}$ & $\mathcal{E}^c_{T_1, T_2}$  \\
         \hline
         Cross-measurement & $\mathcal{E}^c_{p_m}$ & $\mathcal{E}^c_{T_1, T_2}$ \\
         \hline
         Frequency collision & $\hat{U}_{\rm Col}$  & $\mathcal{E}^c_{T_1, T_2}$ \\
         \hline
         $ZZ$-coupling & $\hat{U}_{ZZ}$  & $\mathcal{E}^{a, c}_{T_1, T_2}$ \\
         \hline
    \end{tabular}
    \caption{Error channels used for each of our simulations. $\mathcal{E}^c_{T_1, T_2}$ applies an identity channel to the ancilla, and a phase and amplitude damping channel to the control qubit equivalent to relaxation and dephasing with $T_1 = 345~\mu$s, $T_2 = 280~\mu$s, and a duration $t_m = 0.71~\mu$s, which is meant to be representative of our experimental device. $\mathcal{E}^{a, c}_{T_1, T_2}$ applies the same channel as $\mathcal{E}^c_{T_1, T_2}$ to the control qubit, and a phase and amplitude damping channel to the ancilla qubit with varying $T_1$ (see Fig.~\ref{fig:non-markov}) and $T_2 = T_1/3$. $\mathcal{E}^a_{\rm dep}$ implements identity on the control and the depolarizing channel of Eq.~(\ref{eqn:dep}) on the ancilla. $\hat{U}^c_{\rm Stark}$ and $\mathcal{E}^c_{p_m}$ apply identity on the ancilla, while on the control applying the Stark-shift unitray $Z$-phase error and the cross-measurement error channel of Eq.~(\ref{eqn:col}), respectively. $\hat{U}_{\rm Col}$ implements the two-qubit unitary error of Eq.~(\ref{eqn:col}).}
    \label{tab:sims_models}
\end{table}

For simulations of {\tt mcm-rb} we use 60 random Clifford sequences at each sequence length, and we sweep the error parameter of each model to generate the data points shown in the upper panels of the figures in section \ref{sec:sims}. The calculation of the exact average gate infidelity, $1-\mathcal{F}$, for the two single-qubit control error channels can be done analytically \cite{Emerson_2005}, and the expressions are
\begin{eqnarray}
    &1 - \mathcal{F}_{\rm Stark} = \frac{1}{3}\left(1-\cos(2\phi)\right), \\
    &1 - \mathcal{F}_{\rm CM} = \frac{p_m}{3}.
\end{eqnarray}
To calculate the average gate infidelity on the control qubit for the two-qubit collision error, we must first calculate the effective single-qubit channel this error induces on the control qubit. To do so, following the approach of \cite{Govia20}, we first construct the Choi state of the two-qubit channel
\begin{eqnarray}
    \sigma_{\mathcal{U}_{\rm col}} = \frac{1}{4}\sum_{j,k}\ketbrat{j}{k}\otimes\mathcal{U}_{\rm col}\left(\ketbrat{j}{k}\right),
\end{eqnarray}
where $\{\left|{j}\right>\}$ is an orthonormal basis for the two-qubit Hilbert space and $\mathcal{U}_{\rm col}(\rho) = \hat{U}_{\rm col}\rho\hat{U}^\dagger_{\rm col}$ is the quantum channel representation of the unitary $\hat{U}_{\rm col}$.

For a quantum channel $\mathcal{E}$ that acts on the linear operator space of a Hilbert space $\mathcal{H}$ the Choi state is constructed by acting with the channel $\mathcal{I}\otimes\mathcal{E}$ on a maximally entangled state of the Hilbert space $\mathcal{H} \otimes \mathcal{H}$, where $\mathcal{I}$ is the identity channel. In our case, $\mathcal{H} \otimes \mathcal{H} = \mathcal{H}^a \otimes \mathcal{H}^c \otimes \mathcal{H}^a \otimes \mathcal{H}^c$, where $\mathcal{H}^{c/a}$ is the Hilbert space for the control/ancilla qubit. To calculate the effective channel on the control qubit alone, we perform a partial trace of $\sigma_{\mathcal{U}_{\rm col}}$ over the two copies of the ancilla Hilbert space
\begin{eqnarray}
    \sigma^c_{\mathcal{U}_{\rm col}} = \frac{1}{4}\sum_{j,k}{\rm Tr}_a\left[\ketbrat{j}{k}\right]\otimes{\rm Tr}_a\left[\mathcal{U}_{\rm col}\left(\ketbrat{j}{k}\right)\right].
\end{eqnarray}
For each value of $\Delta/J$ in Fig.~\ref{fig:sim_col} we perform this partial trace numerically to calculate the effective control qubit error channel Choi state, from which we can extract the average gate infidelity.

\section{Full mcm-rb suite data}
\label{app:fulldata}

\begin{figure*}[t!]
    \centering
    \includegraphics[width=\columnwidth]{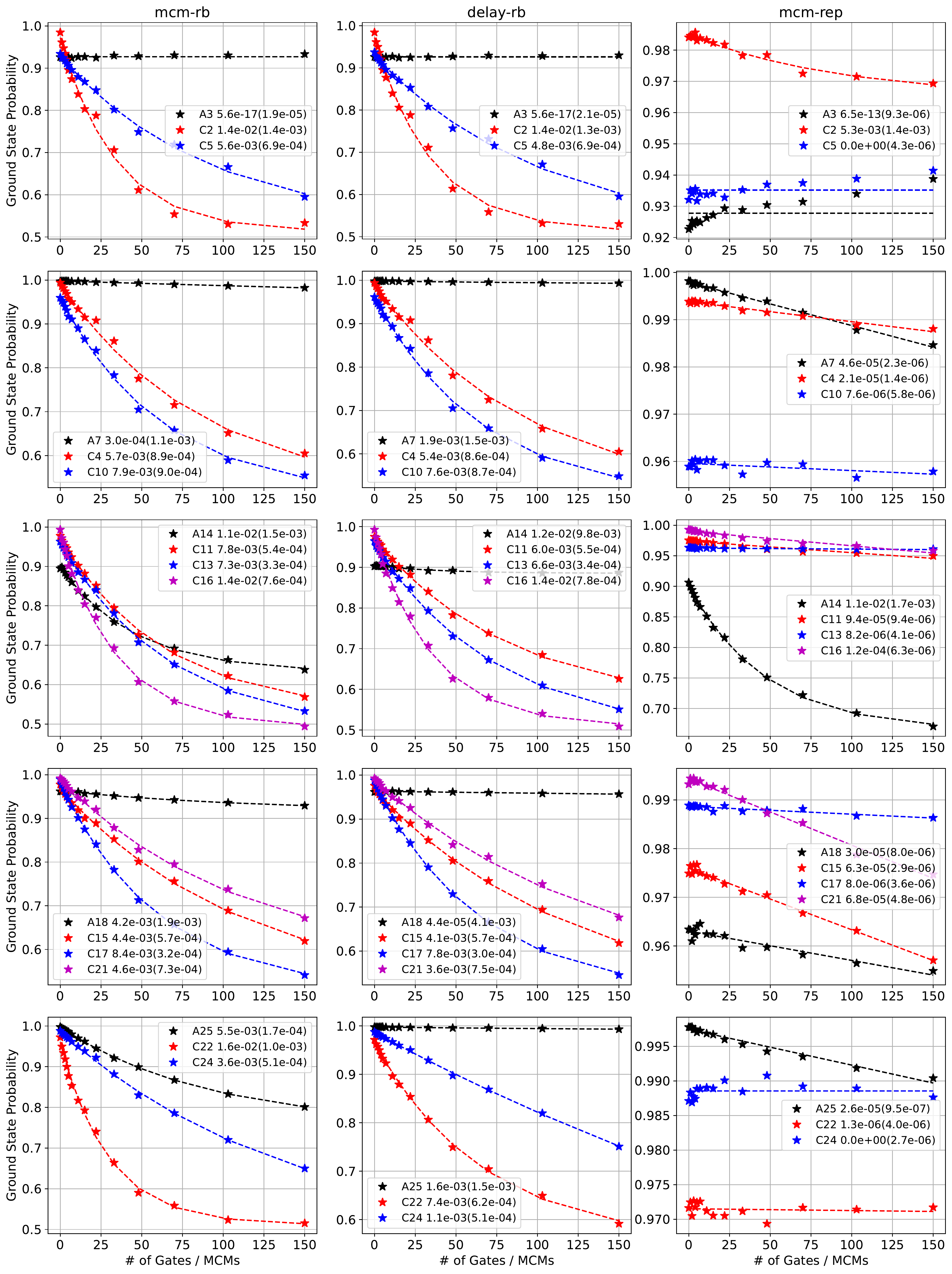}
    \caption{The mcm-rb suite decay curves for the qubits studied in configuration 1 of Fig.~\ref{fig:config} on {\tt ibm\_peekskill}. Each row corresponds to a distinct ancilla-controls group, and each column is one protocol from the mcm-rb suite.}
    \label{fig:config1}
\end{figure*}

\begin{figure*}[t!]
    \centering
    \includegraphics[width=\columnwidth]{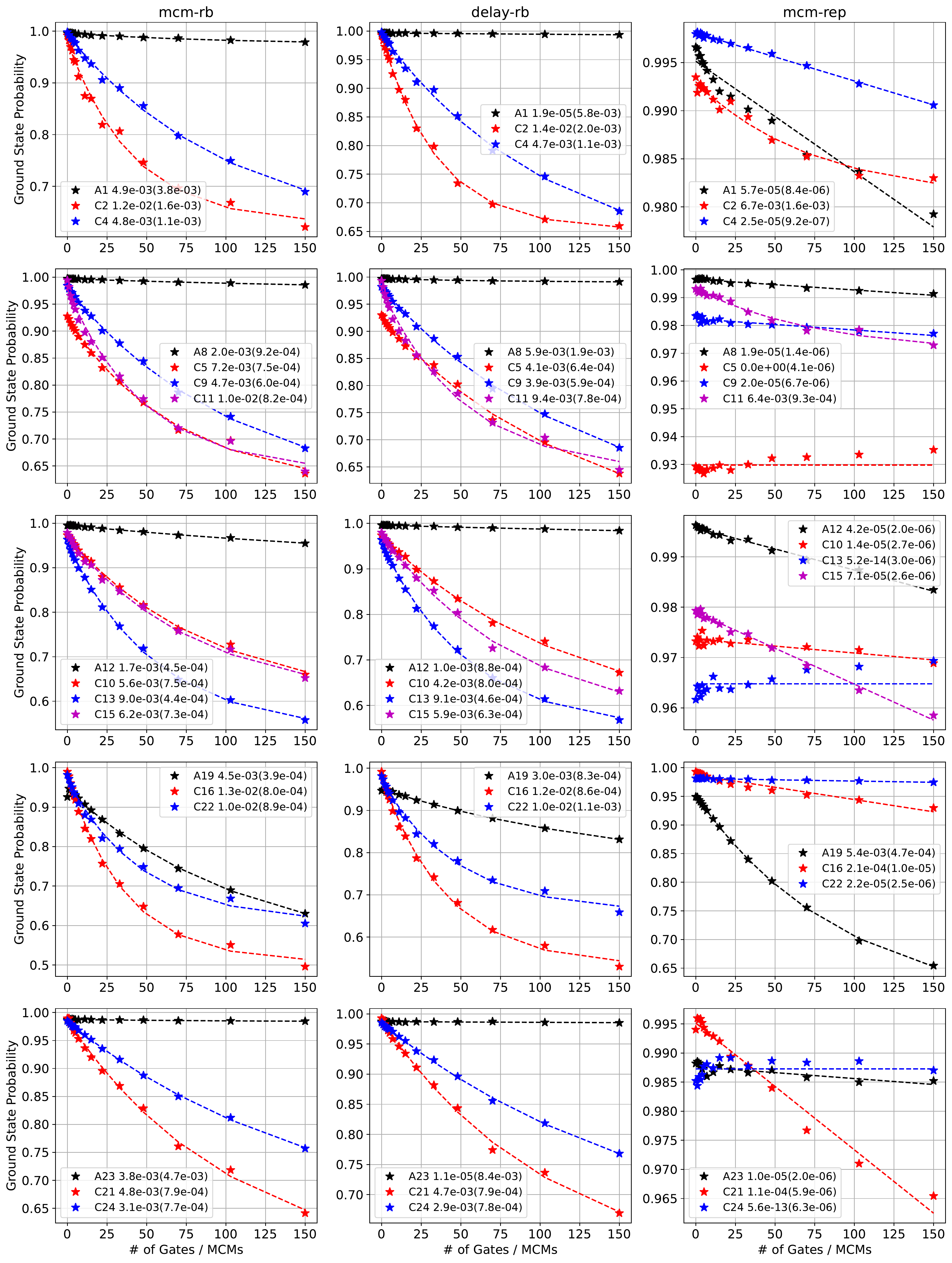}
    \caption{The mcm-rb suite decay curves for the qubits studied in configuration 2 of Fig.~\ref{fig:config} on {\tt ibm\_peekskill}. Each row corresponds to a distinct ancilla-controls group, and each column is one protocol from the mcm-rb suite.}
    \label{fig:config2}
\end{figure*}

The full set of mcm-rb suite decay curves for configuration 1 and configuration 2 on {\tt ibm\_peekskill} are shown in Fig.~\ref{fig:config1} and Fig.~\ref{fig:config2} respectively. For these plots, each row corresponds to a distinct ancilla-controls group shown in Fig.~\ref{fig:config}. Each column is one of the three protocols in the mcm-rb suite.

\section*{References}
\bibliographystyle{iopart-num}
\bibliography{MCMRB_bib}

\providecommand{\newblock}{}
\begin{thebibliography}{10}
\expandafter\ifx\csname url\endcsname\relax
  \def\url#1{{\tt #1}}\fi
\expandafter\ifx\csname urlprefix\endcsname\relax\def\urlprefix{URL }\fi
\providecommand{\eprint}[2][]{\url{#2}}

\bibitem{Egan21}
Egan L, Debroy D~M, Noel C, Risinger A, Zhu D, Biswas D, Newman M, Li M, Brown
  K~R, Cetina M and Monroe C 2021 {\em Nature\/} {\bf 598} 281--286
  \urlprefix\url{https://doi.org/10.1038/s41586-021-03928-y}

\bibitem{Chen21}
Chen Z and {Google Quantum AI} 2021 {\em Nature\/} {\bf 595} 383--387
  \urlprefix\url{https://doi.org/10.1038/s41586-021-03588-y}

\bibitem{Postler2021}
Postler L, Heu{\ss}en S, Pogorelov I, Rispler M, Feldker T, Meth M, Marciniak
  C~D, Stricker R, Ringbauer M, Blatt R, Schindler P, M{\"u}ller M and Monz T
  2022 {\em Nature\/} {\bf 605} 675--680
  \urlprefix\url{https://doi.org/10.1038/s41586-022-04721-1}

\bibitem{Satzinger21}
Satzinger K~J and {Google Quantum AI} 2021 {\em Science\/} {\bf 374} 1237--1241
  \urlprefix\url{https://www.science.org/doi/abs/10.1126/science.abi8378}

\bibitem{Ryan-Anderson21}
Ryan-Anderson C, Bohnet J~G, Lee K, Gresh D, Hankin A, Gaebler J~P, Francois D,
  Chernoguzov A, Lucchetti D, Brown N~C, Gatterman T~M, Halit S~K, Gilmore K,
  Gerber J~A, Neyenhuis B, Hayes D and Stutz R~P 2021 {\em Phys. Rev. X\/} {\bf
  11}(4) 041058
  \urlprefix\url{https://link.aps.org/doi/10.1103/PhysRevX.11.041058}

\bibitem{Abobeih21}
Abobeih M~H, Wang Y, Randall J, Loenen S~J~H, Bradley C~E, Markham M, Twitchen
  D~J, Terhal B~M and Taminiau T~H 2022 {\em Nature\/} {\bf 606} 884--889
  \urlprefix\url{https://doi.org/10.1038/s41586-022-04819-6}

\bibitem{Chen21IBM}
Chen E~H, Yoder T~J, Kim Y, Sundaresan N, Srinivasan S, Li M, C\'orcoles A~D,
  Cross A~W and Takita M 2022 {\em Phys. Rev. Lett.\/} {\bf 128}(11) 110504
  \urlprefix\url{https://link.aps.org/doi/10.1103/PhysRevLett.128.110504}

\bibitem{Krinner21}
Krinner S, Lacroix N, Remm A, Di~Paolo A, Genois E, Leroux C, Hellings C, Lazar
  S, Swiadek F, Herrmann J, Norris G~J, Andersen C~K, M{\"u}ller M, Blais A,
  Eichler C and Wallraff A 2022 {\em Nature\/} {\bf 605} 669--674
  \urlprefix\url{https://doi.org/10.1038/s41586-022-04566-8}

\bibitem{Zhao21}
Zhao Y, Ye Y, Huang H~L, Zhang Y, Wu D, Guan H, Zhu Q, Wei Z, He T, Cao S, Chen
  F, Chung T~H, Deng H, Fan D, Gong M, Guo C, Guo S, Han L, Li N, Li S, Li Y,
  Liang F, Lin J, Qian H, Rong H, Su H, Sun L, Wang S, Wu Y, Xu Y, Ying C, Yu
  J, Zha C, Zhang K, Huo Y~H, Lu C~Y, Peng C~Z, Zhu X and Pan J~W 2021
  Realization of an error-correcting surface code with superconducting qubits
  (\textit{Preprint} \eprint{arXiv:2112.13505})

\bibitem{Sundaresan22}
Sundaresan N, Yoder T~J, Kim Y, Li M, Chen E~H, Harper G, Thorbeck T, Cross
  A~W, Córcoles A~D and Takita M 2022 Matching and maximum likelihood decoding
  of a multi-round subsystem quantum error correction experiment
  (\textit{Preprint} \eprint{arXiv:2203.07205})

\bibitem{Riste13}
Rist{\`e} D, Dukalski M, Watson C~A, de~Lange G, Tiggelman M~J, Blanter Y~M,
  Lehnert K~W, Schouten R~N and DiCarlo L 2013 {\em Nature\/} {\bf 502}
  350--354

\bibitem{Livingston22}
Livingston W~P, Blok M~S, Flurin E, Dressel J, Jordan A~N and Siddiqi I 2022
  {\em Nature Communications\/} {\bf 13} 2307

\bibitem{Luis99}
Luis A and S\'anchez-Soto L~L 1999 {\em Phys. Rev. Lett.\/} {\bf 83}(18)
  3573--3576
  \urlprefix\url{https://link.aps.org/doi/10.1103/PhysRevLett.83.3573}

\bibitem{Fiurasek01}
Fiur\'a\ifmmode~\check{s}\else \v{s}\fi{}ek J 2001 {\em Phys. Rev. A\/} {\bf
  64}(2) 024102
  \urlprefix\url{https://link.aps.org/doi/10.1103/PhysRevA.64.024102}

\bibitem{DAriano04}
D'Ariano G~M, Maccone L and Presti P~L 2004 {\em Phys. Rev. Lett.\/} {\bf
  93}(25) 250407
  \urlprefix\url{https://link.aps.org/doi/10.1103/PhysRevLett.93.250407}

\bibitem{Lundeen09}
Lundeen J~S, Feito A, Coldenstrodt-Ronge H, Pregnell K~L, Silberhorn C, Ralph
  T~C, Eisert J, Plenio M~B and Walmsley I~A 2009 {\em Nature Physics\/} {\bf
  5} 27--30

\bibitem{YChen19}
Chen Y, Farahzad M, Yoo S and Wei T~C 2019 {\em Phys. Rev. A\/} {\bf 100}(5)
  052315 \urlprefix\url{https://link.aps.org/doi/10.1103/PhysRevA.100.052315}

\bibitem{qilgst}
Rudinger K, Ribeill G~J, Govia L~C~G, Ware M, Nielsen E, Young K, Ohki T~A,
  Blume-Kohout R and Proctor T 2022 {\em Phys. Rev. Applied\/} {\bf 17}(1)
  014014
  \urlprefix\url{https://link.aps.org/doi/10.1103/PhysRevApplied.17.014014}

\bibitem{Pereira21}
Pereira L, Garc\'{\i}a-Ripoll J~J and Ramos T 2022 {\em Phys. Rev. Lett.\/}
  {\bf 129}(1) 010402
  \urlprefix\url{https://link.aps.org/doi/10.1103/PhysRevLett.129.010402}

\bibitem{Pereira22}
Pereira L, García-Ripoll J~J and Ramos T 2022 Parallel qnd measurement
  tomography of multi-qubit quantum devices (\textit{Preprint}
  \eprint{arXiv:2204.10336})

\bibitem{Helsen20}
Helsen J, Roth I, Onorati E, Werner A and Eisert J 2022 {\em PRX Quantum\/}
  {\bf 3}(2) 020357
  \urlprefix\url{https://link.aps.org/doi/10.1103/PRXQuantum.3.020357}

\bibitem{Knill08}
Knill E, Leibfried D, Reichle R, Britton J, Blakestad R~B, Jost J~D, Langer C,
  Ozeri R, Seidelin S and Wineland D~J 2008 {\em Phys. Rev. A\/} {\bf 77}(1)
  012307 \urlprefix\url{https://link.aps.org/doi/10.1103/PhysRevA.77.012307}

\bibitem{Magesan11}
Magesan E, Gambetta J~M and Emerson J 2011 {\em Phys. Rev. Lett.\/} {\bf
  106}(18) 180504
  \urlprefix\url{https://link.aps.org/doi/10.1103/PhysRevLett.106.180504}

\bibitem{Magesan12}
Magesan E, Gambetta J~M and Emerson J 2012 {\em Phys. Rev. A\/} {\bf 85}(4)
  042311 \urlprefix\url{https://link.aps.org/doi/10.1103/PhysRevA.85.042311}

\bibitem{Magesan12b}
Magesan E, Gambetta J~M, Johnson B~R, Ryan C~A, Chow J~M, Merkel S~T, da~Silva
  M~P, Keefe G~A, Rothwell M~B, Ohki T~A, Ketchen M~B and Steffen M 2012 {\em
  Phys. Rev. Lett.\/} {\bf 109}(8) 080505
  \urlprefix\url{https://link.aps.org/doi/10.1103/PhysRevLett.109.080505}

\bibitem{Gambetta12}
Gambetta J~M, C\'orcoles A~D, Merkel S~T, Johnson B~R, Smolin J~A, Chow J~M,
  Ryan C~A, Rigetti C, Poletto S, Ohki T~A, Ketchen M~B and Steffen M 2012 {\em
  Phys. Rev. Lett.\/} {\bf 109}(24) 240504
  \urlprefix\url{https://link.aps.org/doi/10.1103/PhysRevLett.109.240504}

\bibitem{Wallman15}
Wallman J, Granade C, Harper R and Flammia S~T 2015 {\em New Journal of
  Physics\/} {\bf 17} 113020

\bibitem{Wallman16}
Wallman J~J, Barnhill M and Emerson J 2016 {\em New Journal of Physics\/} {\bf
  18} 043021

\bibitem{Cross16}
Cross A~W, Magesan E, Bishop L~S, Smolin J~A and Gambetta J~M 2016 {\em npj
  Quantum Information\/} {\bf 2} 16012

\bibitem{Wood18}
Wood C~J and Gambetta J~M 2018 {\em Phys. Rev. A\/} {\bf 97}(3) 032306
  \urlprefix\url{https://link.aps.org/doi/10.1103/PhysRevA.97.032306}

\bibitem{McKay19}
McKay D~C, Sheldon S, Smolin J~A, Chow J~M and Gambetta J~M 2019 {\em Phys.
  Rev. Lett.\/} {\bf 122}(20) 200502
  \urlprefix\url{https://link.aps.org/doi/10.1103/PhysRevLett.122.200502}

\bibitem{Helsen19}
Helsen J, Xue X, Vandersypen L~M~K and Wehner S 2019 {\em npj Quantum
  Information\/} {\bf 5} 71

\bibitem{Proctor19}
Proctor T~J, Carignan-Dugas A, Rudinger K, Nielsen E, Blume-Kohout R and Young
  K 2019 {\em Phys. Rev. Lett.\/} {\bf 123}(3) 030503
  \urlprefix\url{https://link.aps.org/doi/10.1103/PhysRevLett.123.030503}

\bibitem{Erhard19}
Erhard A, Wallman J~J, Postler L, Meth M, Stricker R, Martinez E~A, Schindler
  P, Monz T, Emerson J and Blatt R 2019 {\em Nature Communications\/} {\bf 10}
  5347

\bibitem{McKay20}
McKay D~C, Cross A~W, Wood C~J and Gambetta J~M 2020 Correlated randomized
  benchmarking (\textit{Preprint} \eprint{arXiv:2003.02354})

\bibitem{Proctor21}
Proctor T, Seritan S, Rudinger K, Nielsen E, Blume-Kohout R and Young K 2021
  Scalable randomized benchmarking of quantum computers using mirror circuits
  (\textit{Preprint} \eprint{arXiv:2112.09853})

\bibitem{Morvan21}
Morvan A, Ramasesh V~V, Blok M~S, Kreikebaum J~M, O'Brien K, Chen L, Mitchell
  B~K, Naik R~K, Santiago D~I and Siddiqi I 2021 {\em Phys. Rev. Lett.\/} {\bf
  126}(21) 210504
  \urlprefix\url{https://link.aps.org/doi/10.1103/PhysRevLett.126.210504}

\bibitem{Gaebler21}
Gaebler J~P, Baldwin C~H, Moses S~A, Dreiling J~M, Figgatt C, Foss-Feig M,
  Hayes D and Pino J~M 2021 {\em Phys. Rev. A\/} {\bf 104}(6) 062440
  \urlprefix\url{https://link.aps.org/doi/10.1103/PhysRevA.104.062440}

\bibitem{zenodo}
doi.org/10.5281/zenodo.6815663
  \urlprefix\url{https://doi.org/10.5281/zenodo.6815662}

\bibitem{Braginsky80}
Braginsky V~B, Vorontsov Y~I and Thorne K~S 1980 {\em Science\/} {\bf 209}
  547--557

\bibitem{Gambetta06}
Gambetta J, Blais A, Schuster D~I, Wallraff A, Frunzio L, Majer J, Devoret M~H,
  Girvin S~M and Schoelkopf R~J 2006 {\em Phys. Rev. A\/} {\bf 74}(4) 042318
  \urlprefix\url{https://link.aps.org/doi/10.1103/PhysRevA.74.042318}

\bibitem{Govia15}
Govia L~C~G and Wilhelm F~K 2015 {\em Phys. Rev. Applied\/} {\bf 4}(5) 054001
  \urlprefix\url{https://link.aps.org/doi/10.1103/PhysRevApplied.4.054001}

\bibitem{Pommerening20}
Pommerening J~C and DiVincenzo D~P 2020 {\em Phys. Rev. A\/} {\bf 102}(3)
  032623 \urlprefix\url{https://link.aps.org/doi/10.1103/PhysRevA.102.032623}

\bibitem{Sank16}
Sank D, Chen Z, Khezri M, Kelly J, Barends R, Campbell B, Chen Y, Chiaro B,
  Dunsworth A, Fowler A, Jeffrey E, Lucero E, Megrant A, Mutus J, Neeley M,
  Neill C, O'Malley P~J~J, Quintana C, Roushan P, Vainsencher A, White T,
  Wenner J, Korotkov A~N and Martinis J~M 2016 {\em Phys. Rev. Lett.\/} {\bf
  117}(19) 190503
  \urlprefix\url{https://link.aps.org/doi/10.1103/PhysRevLett.117.190503}

\bibitem{Khezri22}
Khezri M, Opremcak A, Chen Z, Bengtsson A, White T, Naaman O, Acharya R,
  Anderson K, Ansmann M, Arute F, Arya K, Asfaw A, Bardin J~C, Bourassa A,
  Bovaird J, Brill L, Buckley B~B, Buell D~A, Burger T, Burkett B, Bushnell N,
  Campero J, Chiaro B, Collins R, Crook A~L, Curtin B, Demura S, Dunsworth A,
  Erickson C, Fatemi R, Ferreira V~S, Burgos L~F, Forati E, Foxen B, Garcia G,
  Giang W, Giustina M, Gosula R, Dau A~G, Hamilton M~C, Harrington S~D, Heu P,
  Hilton J, Hoffmann M~R, Hong S, Huang T, Huff A, Iveland J, Jeffrey E, Kelly
  J, Kim S, Klimov P~V, Kostritsa F, Kreikebaum J~M, Landhuis D, Laptev P, Laws
  L, Lee K, Lester B~J, Lill A~T, Liu W, Locharla A, Lucero E, Martin S, McEwen
  M, Megrant A, Mi X, Miao K~C, Montazeri S, Morvan A, Neeley M, Neill C,
  Nersisyan A, Ng J~H, Nguyen A, Nguyen M, Potter R, Quintana C, Rocque C,
  Roushan P, Sankaragomathi K, Satzinger K~J, Schuster C, Shearn M~J, Shorter
  A, Shvarts V, Skruzny J, Smith W~C, Sterling G, Szalay M, Thor D, Torres A,
  Woo B~W~K, Yao Z~J, Yeh P, Yoo J, Young G, Zhu N, Zobrist N, Sank D, Korotkov
  A, Chen Y and Smelyanskiy V 2022 Measurement-induced state transitions in a
  superconducting qubit: Within the rotating wave approximation
  (\textit{Preprint} \eprint{arXiv:2212.05097})

\bibitem{Rudinger21}
Rudinger K, Hogle C~W, Naik R~K, Hashim A, Lobser D, Santiago D~I, Grace M~D,
  Nielsen E, Proctor T, Seritan S, Clark S~M, Blume-Kohout R, Siddiqi I and
  Young K~C 2021 {\em PRX Quantum\/} {\bf 2}(4) 040338
  \urlprefix\url{https://link.aps.org/doi/10.1103/PRXQuantum.2.040338}

\bibitem{Qiskit}
 2021 Qiskit: An open-source framework for quantum computing

\bibitem{Ball16}
Ball H, Stace T~M, Flammia S~T and Biercuk M~J 2016 {\em Phys. Rev. A\/} {\bf
  93}(2) 022303
  \urlprefix\url{https://link.aps.org/doi/10.1103/PhysRevA.93.022303}

\bibitem{Beale23}
Beale S~J and Wallman J~J 2023 Randomized compiling for subsystem measurements
  (\textit{Preprint} \eprint{arXiv:2304.06599})

\bibitem{Riste20}
Rist{\`e} D, Govia L~C~G, Donovan B, Fallek S~D, Kalfus W~D, Brink M, Bronn N~T
  and Ohki T~A 2020 {\em npj Quantum Information\/} {\bf 6} 71

\bibitem{Urbanek20}
Urbanek M, Nachman B and de~Jong W~A 2020 {\em Phys. Rev. A\/} {\bf 102}(2)
  022427 \urlprefix\url{https://link.aps.org/doi/10.1103/PhysRevA.102.022427}

\bibitem{Corcoles21}
C\'orcoles A~D, Takita M, Inoue K, Lekuch S, Minev Z~K, Chow J~M and Gambetta
  J~M 2021 {\em Phys. Rev. Lett.\/} {\bf 127}(10) 100501
  \urlprefix\url{https://link.aps.org/doi/10.1103/PhysRevLett.127.100501}

\bibitem{Botelho22}
Botelho L, Glos A, Kundu A, Miszczak J~A, Salehi O and Zimbor\'as Z 2022 {\em
  Phys. Rev. A\/} {\bf 105}(2) 022441
  \urlprefix\url{https://link.aps.org/doi/10.1103/PhysRevA.105.022441}

\bibitem{Piveteau22}
Piveteau C and Sutter D 2022 Circuit knitting with classical communication
  (\textit{Preprint} \eprint{arXiv:2205.00016})

\bibitem{Emerson_2005}
Emerson J, Alicki R and {\.{Z}}yczkowski K 2005 {\em Journal of Optics B:
  Quantum and Semiclassical Optics\/} {\bf 7} S347--S352
  \urlprefix\url{https://doi.org/10.1088/1464-4266/7/10/021}

\bibitem{Govia20}
Govia L~C~G, Ribeill G~J, Rist{\`e} D, Ware M and Krovi H 2020 {\em Nature
  Communications\/} {\bf 11} 1084

\end{thebibliography}

\end{document}